\documentclass[twocolumn]{aastex631}
\newcommand{\msun}{{\rm{M_\odot}}}
\newcommand{\zsun}{{\rm{Z_\odot}}}
\newcommand{\rsun}{{\rm{R_\odot}}}
\newcommand{\rt}{{r_{\rm{t,0}}}}
\newcommand{\rv}{{r_{\rm{v}}}}

\newcommand{\chisq}{{\chi^2_{\nu}}}
\newcommand{\wo}{{W_{\rm{0}}}}
\newcommand{\tdisrupt}{{t_{\rm{disrupt}}}}
\newcommand{\ttrans}{{t_{\rm{trans}}}}
\newcommand{\thalf}{{\rm{T_{1/2,0}}}}

\newcommand{\gyr}{{\rm{Gyr}}}
\newcommand{\myr}{{\rm{Myr}}}

\newcommand{\grida}{\textsf{GRID1}}
\newcommand{\gridb}{{GRID2}}

\newcommand{\pc}{{\rm{pc}}}

\newcommand{\cmc}{{\texttt{CMC}}}
\newcommand{\cosmic}{{\texttt{COSMIC}}}

\newcommand{\nbh}{{N_{\rm{BH}}}}

\newcommand{\fbh}{{f_{\rm{b, h}}}}

\newcommand{\ncluster}{N}
\newcommand{\mcluster}{{M_{\rm{cluster}}}}

\usepackage{graphicx}	
\usepackage{amsmath}	
\usepackage{amssymb}	
\usepackage{enumitem}

\begin{document}

\shorttitle{Dynamical evolution of young star clusters}

\title{From Young Massive Clusters to Old Globular Clusters: Density Profile Evolution and IMBH Formation}

\shortauthors{Sharma \& Rodriguez}


\author[0000-0002-6675-0424]{Kuldeep Sharma}
\affiliation{McWilliams Center for Cosmology\\
Carnegie Mellon University\\
5000 Forbes Ave, Pittsburgh, PA 15213, USA}
\affiliation{Department of Physics and Astronomy\\
    University of North Carolina at Chapel Hill\\
    120 E. Cameron Ave, Chapel Hill, NC, 27599, USA}
\email{kul.shar90@gmail.com}

\author[0000-0003-4175-8881]{Carl L. Rodriguez}
\affiliation{Department of Physics and Astronomy\\
    University of North Carolina at Chapel Hill\\
    120 E. Cameron Ave, Chapel Hill, NC, 27599, USA}


\begin{abstract}
The surface brightness profiles of globular clusters are conventionally described with the well-known King profile. However, observations of young massive clusters (YMCs) in the local Universe suggest that they are better fit by simple models with flat central cores and simple power-law densities in their outer regions (such as the Elson-Fall-Freeman, or EFF, profile).   Depending on their initial central density, YMCs may also facilitate large numbers of stellar collisions, potentially creating very massive stars that will directly collapse to intermediate-mass black holes (IMBHs).  Using Monte Carlo $N$-body models of YMCs, we show that EFF-profile clusters transform to Wilson or King profiles through natural dynamical evolution, but that their final $W_0$ parameters do not strongly correlate to their initial concentrations.  In the densest YMCs, runaway stellar mergers can produce stars that collapse into IMBHs, with their final masses depending on the treatment of the giant star envelopes during collisions. If a common-envelope prescription is assumed, where the envelope is partially or entirely lost, stars form with masses up to $824\,\msun$, collapsing into IMBHs of $232\,\msun$. Alternatively, if no mass loss is assumed, stars as massive as $4000\,\msun$ can form, collapsing into IMBHs of $\sim 4000\,\msun$.  In doing so, these runaway collisions also deplete the clusters of their primordial massive stars, reducing the number of stellar-mass BHs by as much as $\sim$ 40\%.  This depletion will accelerate the core collapse, suggesting that the process of IMBH formation itself may produce the high densities observed in some core-collapsed clusters.

\end{abstract}




\section{Introduction}
Pair-instability supernovae (PISN) and pulsational pair-instability supernovae (PPSN) suggest that it should not be possible to form black holes (BHs), through the evolution of single stars, in the mass range $\sim 50-120 \, \msun$, termed the ``upper mass gap'' \citep[e.g.,][]{Spera2017, Belczynski2016, Woosley2007}. At the onset of carbon burning in an evolved star with a helium core mass in the $45-135 \, \msun$ range, electron-positron pairs are produced \citep{Barkat1967}. The pair production depletes the energy of the star, reducing the internal radiation pressure. This reduction in internal pressure leads to a partial collapse of the core, which accelerates nuclear burning of heavy elements, culminating in a runaway thermonuclear explosion. For the stars with helium core mass in the range $\sim 45-65 \, \msun$, the pair-instability leads to PPSN: violent pulsations that reduce the mass and entropy of helium and other heavy elements until the pulsations damp out \citep[e.g.,][]{Heger2002, Woosley2007, Woosley2017}. If the helium core mass is larger than $65 \, \msun$ then the star undergoes PISN and the thermonuclear explosion completely destroys the star. 

However, the gravitational wave event GW190521, resulting from the merger of two BHs of masses $66 \, \msun$ and  $85 \, \msun$ was the first observational evidence of BHs existing in the ``upper mass gap''. BHs with masses in the range of $10^2 - 10^5 \, \msun$ are called intermediate-mass black holes \citep[IMBHs, see][for a review]{Greene2020}.
The Gravitational-Wave Transient Catalog 3 \citep[GWTC-3,][]{GWTC3_2021} compiled by including all the gravitational wave events from the first three observing runs of the Advanced Laser Interferometer Gravitational-Wave Observatory  \citep[LIGO,][]{Aasi2015} and Advanced Virgo \citep{Acernese2014}, lists fifteen Binary Black Hole (BBH) merger events with at least one component mass higher than $45 \, \msun$, three events with both component masses above $45 \, \msun$, and six with at least one BH more massive than $60 \, \msun$  \citep{Abbott2021}. It is imperative to understand the formation scenarios of BHs with a mass in or above the mass gap. 

There are several proposed mechanisms for the formation of BHs with masses $\sim 50 - 10^5 \, \msun$. Stars in the early universe (Pop III stars) were extremely metal-poor and mostly composed of hydrogen \citep[e.g.,][and references therein]{Madau2001, Bromm2004} . The inability of molecular hydrogen to efficiently cool down would have led to Pop III stars being quite massive \citep[see][for a review]{Karlsson2013}. The first formation scenario involves the direct collapse of these massive stars, resulting in massive BHs \citep{Fryer2001}. Unfortunately, detecting Pop III stars, whether isolated or in clusters, appears to be beyond the reach of even the James Webb Space Telescope \citep[JWST,][]{Rydeberg2013}. Second, gravitational instabilities during the inflationary era could have formed BHs of $\sim 1 - 10^3 \, \msun$ \citep{Carr2016}, and BHs of  $\sim 10^3-10^5 \, \msun$ in the early universe from collapsing gas clouds without going through all the stages of stellar evolution \citep[e.g.,][]{Loeb1994, Bromm2003}. These BHs can serve as seeds for the supermassive BHs seen at the centers of present-day galaxies. Third, gravitational runaway mergers in high-density central environments of star clusters can produce BHs of mass $10^2 - 10^4 \, \msun$. Gravitational runaways can be divided into two categories based on timescale. The slow scenario, unfolding over a timescale of $100\, \myr$ to a $\gyr$, involves hierarchical mergers of smaller BHs, and can produce IMBHs of mass $10^2 - 10^3 \, \msun$  \citep[e.g.,][]{Miller2002, Mckernan2012, Giersz2015, Carl2019, Fragione2020, Kroupa2020, Fragione2020_1, Fragione2022, Rizzuto2022}.

The fast scenario involves collisional runaway mergers of massive stars in dense star clusters and unfolds on a short timescale of tens of Myrs early in the cluster’s lifetime  \citep{Zwart2002}. A star can collide and merge with another star, grow its mass and physical size, and then go on to merge with other stars repeatedly in a runaway fashion. The resultant product star can become very massive, may not undergo PISN, and can collapse to form either mass gap BHs or IMBHs \citep[e.g.,][]{Zwart2004, Giersz2015, Banerjee2020, Kremer2020, Banerjee2021, Ugo2021, Rizzuto2021, Rizzuto2022, Ballone2022}.

Observational evidence suggests that the majority of stars form in clustered environments, particularly so for massive stars \citep[][and references therein]{Liu2021, Zwart2010}. 
Given that massive stars play a crucial role in the collisional runaway mechanism of IMBH formation, it is natural to investigate the dynamical evolution of star clusters to understand collisional runaways. Various types of star clusters exist, including globular clusters (GCs), open clusters, nuclear star clusters, and young massive clusters (YMCs). YMCs typically have a mass $>10^4\,\msun$ and age $<100\,$Myr \citep{Zwart2010}, and it  is particularly interesting to study their evolution for two reasons. Firstly, they are considered the most common birthplaces of massive stars, characterized by high central densities and young dynamical states, making them ideal candidates for the stellar collisional runaway channel of IMBH formation \citep{Zwart2004, Rizzuto2021}. Secondly, YMCs are believed to be progenitors of present-day GCs \citep{Kruijssen2012}. By studying the evolution of YMCs, we can gain insights into the processes that lead to the formation of GCs and their subsequent dynamical evolution over cosmic timescales.

\cite{Kremer2020} explored the high-mass star cluster regime ($10^5 - 10^6 \, \msun$), which is representative of the Milky-Way GC population \citep{Harris1996}, using a suite of 68 clusters with an initial \cite{King1966} density profile simulated using \texttt{Cluster Monte Carlo} (\cmc) code. They found that approximately 20\% of all BH progenitors undergo at least one stellar collision prior to collapse and about 1\% of all BHs reside in the upper mass gap. Many of the BBHs produce second-generation BHs through mergers within the cluster. \citet{Gonzalez2021} investigated the effect of the binary fraction of stars more massive than $15 \, \msun$  on the formation of IMBHs in YMCs. Their suite of simulations, with a fixed initial cluster mass of $4.7\times10^5 \, \msun$, metallicity of $0.1 \, \zsun$ and a King density profile, demonstrated that YMCs with larger high-mass binary fractions are more efficient in forming IMBHs.

\citet{Gonzalez2021} used a King profile for their YMCs. However, observations strongly suggest that YMCs have extended outer halos and their surface brightness profiles are often better fitted with Elson-Fall-Freeman (EFF) profiles \citep[e.g.,][]{Elson1987, Larsen1999, MG2003LMC, McLaughlin2005, Bastian2013} than with King profiles. At fixed virial radii, EFF profile clusters can have high central densities, making them dynamically more active and fertile for collisional runaways. 

In this paper, we study the effect of high-mass binary fraction, initial cluster size, cluster mass, and giant star collision prescription on the short-term evolution of YMCs with an initial EFF profile. We discuss the evolution of the cluster profile from an EFF to a King profile and emphasize the formation of IMBHs through stellar runaway collisions.  This paper is organized as follows: In Section~\ref{sect:sims}, we provide a brief discussion of the \cmc\ code and detail the initial conditions for cluster simulations. Section~\ref{sect:fitprofs} outlines the fitting procedure for various density profiles to the cluster surface brightness.  In Section~\ref{sect:proftrans}, we discuss the evolution of the cluster density profile using the data from simulations. Section~\ref{sect:runaways} describes important modifications made to the \cmc\ code to handle massive star collisions and presents results from the simulation grid used to study IMBH formation through stellar collisional runaways. Finally, we summarize our findings and discuss future prospects in Section~\ref{sect:discussion}.

\section{Cluster simulations}\label{sect:sims}
We use the H\'enon style Monte Carlo code \cmc\ to simulate the evolution of star clusters. \cmc\ includes up-to-date prescriptions for  mass loss rate of massive stars, compact object formation, strong 3 and 4 body encounters using \texttt{Fewbody} \citep{Fregeau2004}, gravitational-wave emission, and stellar/binary evolution prescriptions with the population synthesis code \cosmic\ \citep{Breivik2020}. We use the delayed compact object formation prescription of \cite{Fryer2012}. Please refer to sections 2 and 3 of \cite{Rodriguez2022} for a discussion on various prescriptions for solving dynamical interactions within a star cluster and an overview of \cmc\ package. 

\citet{Elson1987} found that the surface brightness profiles of young star clusters are better modeled by power law fits compared to King profiles which are used to fit old evolved GCs in the MW. In EFF profiles, the 2D surface density profile is given by - 
\begin{equation}\label{eqn:elson2d}
\mu(r) = \mu_{\rm 0}\,\left(1 + \frac{R^2}{a^2}\right)^{-\frac{\gamma}{2}}
\end{equation}
where $\mu_{\rm 0}$ is the maximum (i.e. central) surface density, and R is the projected distance from the center of the cluster. The corresponding three-dimensional mass density $\rho(r)$ at a radius $r$ is given by - 

\begin{equation}\label{eqn:elson3d}
\rho(r) = \rho_{\rm 0}\,\left(1 + \frac{r^2}{a^2}\right)^{-\frac{\gamma  +1}{2}}
\end{equation}
where $\rho_{\rm 0}$ is the mass density at the center of the cluster, $a$ is the scale radius, and $\gamma$ is the power law slope of the profile.

\begin{figure*}
\centering
 \includegraphics{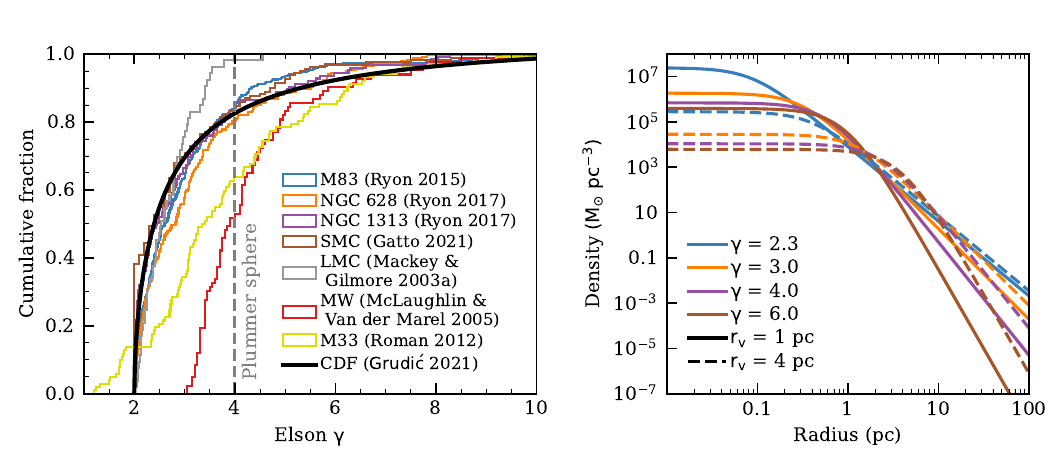}
 \caption{\textit{Left panel :} shows observational data for EFF profile slopes of star clusters from the literature, along with suggested universal CDF of EFF slopes from \citep{Grudic2021}. fitted using the observational data of young star clusters in the age range $10^6 - 10^8$ yr. The histograms show the EFF profile slopes from the observed surface brightness of star clusters in M83 (age $< 3 \times 10^6$ yr) from \cite{Ryon2015}, NGC 628 and NGC 1313 (age $< 2 \times 10^6$ yr) from \cite{Ryon2017}, M33 (age :  $10^7 - 10^9$ yr) from \cite{Roman2012}, the Large Magellanic Cloud (age : $10^6 - 10^{10}$ yr) from \cite{MG2003LMC}, the Small Magellanic Cloud (age : $10^6 - 10^{10}$ yr) from \cite{Gatto2021}, and the MW (age $ > 7 \times 10^9$ yr) from \cite{McLaughlin2005}. The CDF curve agrees well with the histograms of young star clusters. This figure is largely a reproduction of figure 2 of \cite{Grudic2018} with additional data for M33 and MW clusters. \textit{Right panel :} shows the density of EFF profiles with different Elson $\gamma$ and virial radii. For a fixed virial radius of the cluster, the central density $\rho_{\rm 0}$ increases as Elson $\gamma$ decreases. Furthermore, for a given Elson $\gamma$ the central density $\rho_{\rm 0}$ decreases as the virial radius of the cluster increases. Therefore, star clusters with small Elson $\gamma$ and small virial radius are the most centrally dense and likely to facilitate IMBH formation through runaway stellar collisions.}
 \label{fig3a}
\end{figure*}

The left panel of Figure~\ref{fig3a} shows the distribution of the EFF profile slopes for young and old star clusters, as reported in the literature. Approximately 92\% of young star clusters have an EFF profile slope between 2.0 and 6.0 with a median of 2.4. It is noteworthy that the distribution of EFF profile slopes differs between the older GC populations observed in M33 and the Milky Way compared to the profiles of young star clusters.

We create two separate grids of simulations with realistic stellar evolution and a metallicity of 0.002 (equivalent to $0.1 \, \zsun$ assuming a solar metallicity of 0.02). The first grid (\grida) of simulations is to study the evolution of the star cluster density profile from the EFF profile to an evolved GC profile like those in the MW. The second grid (\gridb) is to study the effect of various physical parameters and initial conditions on runaway collisions and massive BH formation. In both grids the masses of single stars and primary stars (within a binary) are sampled from \cite{kroupa2001} initial mass function in the mass range of 
$0.08 - 150 \, \msun$.

\grida\ consists of initial conditions with EFF profile slopes $\gamma \in \{2.3, 3.0, 4.0, 6.0\}$ to capture the 92\% of the range of $\gamma$ for young clusters. These clusters have cluster tidal radii, $\rt \in \{5, 10, 20, 50, 100\} \, \pc$, and initial virial radii, $\rv \in \{1, 4\} \, \pc$. For a cluster with mass $ M_{\rm C}$ embedded in a galaxy with mass $ M_{\rm G}$ at a distance of $R_{\rm G}$, the tidal radius $\rt$ is given as 

\begin{equation}\label{eqn:rtid}
\rt = \left(\frac{M_{\rm C}}{3 M_{\rm G}}\right)^{1/3}  R_{\rm G} 
\end{equation}

All clusters have an initial cluster mass of $\mcluster \simeq 3\times10^5 \, \msun$ corresponding to $\ncluster = 5 \times 10^5$ stars and binaries, and a 10\% initial binary fraction. This grid uses the default prescriptions in \cmc\ for dynamics and stellar evolution.

\gridb\ consists of initial conditions with EFF profile slopes $\gamma \in \{2.3, 3.0, 4.0, 6.0\}$; initial cluster mass of $\mcluster \simeq 6\times10^5 \, \msun$ and $\mcluster \simeq 3\times10^5 \, \msun$ corresponding to $\ncluster = 10^6$ and $\ncluster = 5\times10^5$ stars and binaries, respectively; initial virial radii, $\rv \in \{1, 4\} \, \pc$; and initial binary fraction for high-mass binaries ($M > 15 \, \msun$) of 5\% and 100\%. Binary fraction for all the stars less massive than $15 \, \msun$ is kept 5\% in all the simulations. Each of these combinations is repeated for three different realizations (using different random number seeds) for two different prescriptions for stellar collisions involving giant stars (discussed below). This results in 288 star-cluster simulations. Each simulated cluster experiences a tidal force corresponding to being on a 8 kpc circular orbit inside the Milky Way. 

The collision process of two main-sequence (MS) stars is modeled using the ``sticky-sphere'' approximation, where no mass is lost during the collision, and the mass of the merger product is the sum of the masses of the colliding stars. The velocity dispersion of stars in globular clusters is on the order of a few tens of km/s, which is substantially lower than the escape velocity from the surface of a MS star. For example, a star with a mass of $M=1\,\msun$  and a radius of $R=1\,\rsun$ has an escape velocity given by $\sqrt{2GM/R} \simeq 617.7\,$km/s. Consequently, collisional trajectories in such environments can be effectively approximated as parabolic, leading to relatively gentle mergers in which the mass loss does not exceed approximately 8\% of the total system mass \citep{Lombardi2002}. Therefore, the assumption of ``sticky-sphere'' collisions for MS-MS interactions is well justified.

In a collision of two MS stars of masses $M_1$ and $M_2$, the stellar age of the new product MS star is given by 
\begin{equation}\label{eqn:rejuv}
t_3 = f_{\rm rejuv} \frac{t_{\rm MS3}}{M_3} \left( \frac{M_1t_1}{t_{\rm MS1}} + \frac{M_2t_2}{t_{\rm MS2}} \right)  
\end{equation}
where $t_{\rm MS1}$, $t_{\rm MS2}$, and $t_{\rm MS3}$ are the MS lifetimes of the colliding stars and the collision product, respectively. The mass of the merger product is $M_3 = M_1\, + \, M_2$; $t_1$ and $t_2$ are the stellar ages of the colliding MS stars at the time of collision. The factor $f_{\rm rejuv}$ determines the amount of rejuvenation in the collision product through  mixing of material of the colliding stars. We use the default value of $f_{\rm rejuv} = 1$ as suggested by \cite{Breivik2020}.

However, when at least one of the colliding stars is a giant the outcome is less clear. In \cosmic\, the merger is treated as a common-envelope evolution, where the cores of the stars orbit within the loosely bound common envelope. Common envelope evolution is modeled through the standard $\alpha \lambda$ prescription \cite[see \S 3.2 of][]{Breivik2020} where $\lambda$ is a factor that determines the binding energy of the envelope to its stellar core, while $\alpha$ is the efficiency factor for injecting orbital energy into the envelope. We use a value of $\alpha=1$ and a variable $\lambda$ that depends on the evolutionary state of the star. We adopt two different prescriptions for the collisions involving a giant star: 1) we make one attempt to merge the stars through common envelope prescription.  
If the stars do not merge in the first attempt, then we force them to merge (referred to as CE2, hereafter). This is the default behavior of \cmc\ for such collisions. In this scenario some or all of the envelope of the giant star is lost. 2) stars are forced to merge in the first attempt itself (referred to as CE0, hereafter). For the merger product, the total mass is again computed as  $M_3 = M_1\, + \, M_2$ and the core mass is computed as $M_{\rm c,3} = M_{\rm c,1}\, + \, M_{\rm c,2}$, where  $M_{\rm c,1}$ and $M_{\rm c,2}$ are the core masses of the colliding stars. This scenario  mimics a sticky sphere collision.  

By default in \cmc\, the collision of a MS star or giant and a BH results in complete disruption of the star. While this prescription is reasonable for near-equal mass encounters, it likely overpredicts the disruption of massive stars and giants by small stellar-mass BHs (where at most collision speeds encountered in GCs, the BH would pass through the envelope of a giant with minimal disruption).  To that end, we modify \cmc\ such that whenever a compact object with mass $M_{\rm comp}$ collides with a giant star with mass $M_* > 100 \, \msun$ and $M_{\rm comp} < M_*/2$, both colliding objects remain unaffected. 

\begin{deluxetable*}{@{\extracolsep{\fill}}lccccc|lccccr@{}}
\tablecaption{Relevant parameters for profile transition in simulations with $\ncluster = 5\times10^5$ objects. \label{tab:transitions}}
\tabletypesize{\small}
\tablehead{
 \colhead{$\gamma$} & \colhead{$\rt$} & \colhead{$\tdisrupt$} &
 \multicolumn{2}{c}{$\ttrans (\thalf)$} & \colhead{King $\wo$} & 
 \colhead{$\gamma$} & \colhead{$\rt$} & \colhead{$\tdisrupt$}  &
 \multicolumn{2}{c}{$\ttrans (\thalf)$} & \colhead{King $\wo$}\\
 \cline{4-5} \cline{10-11}
 \colhead{} & \colhead{($\pc$)} & \colhead{$\myr$} &
 \colhead{ Wilson} & \colhead{King} & \colhead{} & 
 \colhead{} & \colhead{($\pc$)} & \colhead{$\myr$} &
 \colhead{ Wilson} & \colhead{King} & \colhead{}\\
 \hline
 \multicolumn{6}{c|}{\rule[10pt]{0pt}{0pt} $\rv = 1\,\pc$} & \multicolumn{6}{c}{\rule[10pt]{0pt}{0pt} $\rv = 4\,\pc$}
 }
\startdata
2.3 & 5.0 & 90 & - & 0.009 & 0.26 & 2.3 & 5.0 & 916 & 0.009  & 0.035 & 0.18 \\
2.3 & 10.0 & 691 & 0.088 & 0.264 & 3.59 & 2.3 & 10.0 & 7532 & 0.088 & 0.351 & 2.84\\
2.3 & 20.0 & 3704 & 0.088 & 1.757 & (7.19)\,9.70 & 2.3 & 20.0 & $> 13800$ & 0.176 & 2.724 & 5.63 \\
2.3 & 50.0 & $> 13800$ & 2.636 & 27.675 & (8.18)\,16.00 & 2.3 & 50.0 & $> 13800$ & 1.933 & - & -\\
2.3 & 100.0 & $> 13800$ & 13.618 & - & - & 2.3 & 100.0 & $> 13800$ & 8.083 & - & - \\[1ex]
\hline
 & & & & & & & & & & &\\[-3ex]
3.0 & 5.0 & 76 & - & 0.012 & 0.10 & 3.0 & 5.0 & 758 & - & 0.012 & 0.12\\
3.0 & 10.0 & 616 & 0.118 & 1.065 & 0.33 & 3.0 & 10.0 & 6572 & 0.118 & 1.184 & 0.13 \\
3.0 & 20.0 & 5119 & 0.947 & 5.682 & (6.34)\,9.54 & 3.0 & 20.0 & $> 13800$ & 0.474 & 6.511 & 4.87\\
3.0 & 50.0 & $> 13800$ & 7.694 & 58.003 & 8.36 & 3.0 & 50.0 & $> 13800$ & 0.355 & - & -\\
3.0 & 100.0 & $> 13800$ & 27.818 & - & - & 3.0 & 100.0 & $> 13800$ & - & - & - \\[1ex]
\hline
 & & & & & & & & & & &\\[-3ex]
4.0 & 5.0 & 73 & 0.011 & 0.043 & 0.13 & 4.0 & 5.0 & 554 & 0.011 & 0.076 & 0.14\\
4.0 & 10.0 & 598 & 0.215 & 1.291 & 0.15 & 4.0 & 10.0 & 5289 & 0.215 & 2.043 & 0.14\\
4.0 & 20.0 & 4565 & 2.581 & 9.679 & 4.15 & 4.0 & 20.0 & 503 & - & - & -\\
4.0 & 50.0 & $> 13800$ & 13.443 & 70.977 & 7.70 & 4.0 & 50.0 & $> 13800$ & - & - & -\\
4.0 & 100.0 & $> 13800$ & 41.403 & - & - & 4.0 & 100.0 & $> 13800$ & - & - & -\\[1ex]
\hline
 & & & & & & & & & & &\\[-3ex]
6.0 & 5.0 & 71 & 0.010 & 0.109 & 0.10 & 6.0 & 5.0 & 479 & 0.010 & 0.070 & 0.10\\
6.0 & 10.0 & 630 & 0.297 & 1.487 & 0.12 & 6.0 & 10.0 & 5667 & 0.397 & 1.785 & 0.16\\
6.0 & 20.0 & 5919 & 3.173 & 10.510 & (6.88)\,9.93 & 6.0 & 20.0 & 5038 & - & - & -\\
6.0 & 50.0 & $> 13800$ & 17.351 & 68.909 & 8.12 & 6.0 & 50.0 & $> 13800$ & - & - & - \\
6.0 & 100.0 & $> 13800$ & 40.156 & - & - & 6.0 & 100.0 & $> 13800$ & - & - & -\\
\enddata

\tablecomments{The left and right halves of the table are for cluster simulations with initial virial radii of $1 \, \pc$ and $4 \, \pc$, respectively. Each half has 6 columns: columns (1-2) are the initial EFF profile slope ($\gamma$) and tidal radius ($\rt$) of the cluster, and column 3 is the disruption time ($\tdisrupt$) of the cluster in Myr. Columns  (4-5) show the transition time ($\ttrans$) for EFF $\rightarrow$ Wilson and EFF $\rightarrow$ King, respectively. The transition time, \ensuremath{\mathbf{\ttrans}}, is reported in units of the cluster's half-mass relaxation time at t=0 (i.e.,  $\thalf$). There are four possible scenarios for the values in columns (4-5) : (- , -) means that the cluster does not transition to either Wilson or King profiles and remains an EFF profile, ($t$, -) shows that the cluster transitions to a Wilson profile at time $t$ and remains a Wilson cluster thereafter. (-, $t$) represents a cluster which transitions to a King profile at time t without transitioning to a Wilson profile. ($t_{\rm 0}, t_{\rm 1}$) shows a typical cluster that first transitions to a Wilson profile at time $t_{\rm 0}$ and then to a King profile at time $t_{\rm 1}$. Finally, column 6 shows the King parameter $\wo$ for the cluster of the best-fit model at the end of the simulation is King. If a cluster does not transition to a King-profile then column 6 is left empty. Some clusters undergo a core collapse and the value of $\wo$ given in bracket is right before the core collapse for such models. Tidally disrupted clusters have $\wo \sim 0.1$.}
\end{deluxetable*}

\vspace*{-0.8cm}
\section{Fitting the surface brightness profiles}\label{sect:fitprofs}
  
It is challenging to determine the three-dimensional positions of stars in clusters through observations without accurate distance information for individual stars. Therefore, it is common for observers to fit projected 2D profiles of theoretical models to the binned surface brightness data of star clusters. 

Our simulations incorporate realistic stellar evolution, providing luminosities and effective temperatures for stars at each snapshot. We use \texttt{cmctoolkit} \citep{Rui2021} to generate the V-band surface brightness profile of star clusters at each snapshot. For each simulation snapshot, we compute the surface brightness and its corresponding Poisson uncertainties using 60 radial bins, spanning from $10^{-2}\,\pc$ to the cluster’s outermost object. \texttt{cmctoolkit} uses the individual stellar luminosities and statistical orbit averaging to generate surface brightness profiles from a given snapshot. 

We fit theoretical profiles to the generated surface brightness profile at each snapshot for the clusters in \grida. First, we project the 3D density $\rho(r)$ of the theoretical model onto a two-dimensional plane by integrating along the line of sight \citep[e.g.,][]{Binney2008}

\begin{equation}\label{eqn:sigma}
\Sigma (R) = 2 \int_R^{\infty} \rho(r) \, \frac{r}{\sqrt{(r^2 - R^2)}} dr
\end{equation}
\noindent Here, $R$ is the projected radial distance from the center of the cluster. Assuming that light follows mass for stars (as our theoretical models are single-mass models), we can write the surface brightness as 

\begin{equation}\label{eqn:mu}
 \mu(R) = \mu_{\rm 0} - 2.5 \log_{\rm 10}(\Sigma (R))
\end{equation}

At each snapshot, we fit the density profiles of \cite{King1966}, \cite{Wilson1975}, and EFF \citep{Elson1987} models to the surface brightness profile of star clusters. For each of these isotropic density models, we use reduced chi-square ($\chisq$) minimization to find the best-fit parameters at each snapshot, and to compare the relative goodness of fits between the three model families.  

The \cite{King1966} distribution function is a single-mass lowered isothermal distribution function with a finite size defined by 
\begin{equation}\label{eqn:kingdf}
    f (\mathcal{E}) = \begin{cases} \rho_{\rm 1} (2\pi\sigma^2)^{-3/2} (e^{\mathcal{E}/\sigma^2} - 1) & \mathcal{E} >0\\ 
                 0 & \mathcal{E}\le 0 \end{cases}
\end{equation}

\noindent Here, $\mathcal{E} = -\Phi - \frac{1}{2}v^2 $ represents the relative energy of a star with velocity $v$ in a cluster with potential $\Phi$. At the tidal boundary of the system, $\mathcal{E}$ goes to 0. The parameters $\rho_{\rm 1}$ and $\sigma$ are normalization factors, with $\sigma$ not to be confused with the actual velocity dispersion of stars in the cluster. It represents a theoretical parameter that only coincides with the velocity dispersion in any given direction under the idealized assumption of an infinite tidal boundary.

The King model does not have an analytic expression for a density profile, and we have to numerically integrate an ODE to find the density and enclosed mass at a given radial position. A dimensionless central potential, $\wo = \sqrt{\frac{-\Phi(0)}{\sigma^2}}$, uniquely sets the density profile of the system. We can find the unnormalised density, $\rho (r)/ \rho_{\rm 1}$, for a given King model defined by the parameter $\wo$. The physical size of the system is determined by the King radius $r_{\rm 0} = \frac{9 \sigma^2}{4\pi G \rho_{\rm 0}}$, where $\rho_{\rm 0}$ is the central mass density, and $G$ is the gravitational constant.  Using $\rho(r)$, we can find the surface brightness profile of a cluster with a given $\wo$ using equations \ref{eqn:sigma} and \ref{eqn:mu}. We used flat priors for $\wo$ in the range [0.1, 16] while fitting the surface brightness profiles due to numerical problems for $\wo < 0.1$. For this reason, most of the disrupted clusters have $\wo \sim 0.1$ in Table~\ref{tab:transitions}. 

The Wilson profile fits the GC surface brightness data better than (or equally as well as) the king models, as reported by \cite{McLaughlin2005}. The single-mass model distribution function for the Wilson profile is given by
\vspace{-0.2cm}
\begin{equation}\label{eqn:wilsondf}
    f (\mathcal{E}) = \begin{cases} \rho_{\rm 1} (2\pi\sigma^2)^{-3/2} (e^{\mathcal{E}/\sigma^2} - 1 - \mathcal{E}/\sigma^2) & \mathcal{E} >0\\ 
                 0 & \mathcal{E}\le 0 \end{cases}
\end{equation}

Similar to the King model, the parameter $\wo$ sets the surface brightness profile of the system and the King radius ($r_{\rm 0}$) sets the scale of the system.

As evident from equation \ref{eqn:elson2d}, for the EFF model, the power law exponent ($\gamma$) sets the surface brightness profile of the system, while the scale radius ($a$) determines the physical size. 

\begin{figure*}[!htbp]
 \centering
 \includegraphics[width=\textwidth, trim={0in 0in 0in 0in}, clip=True]{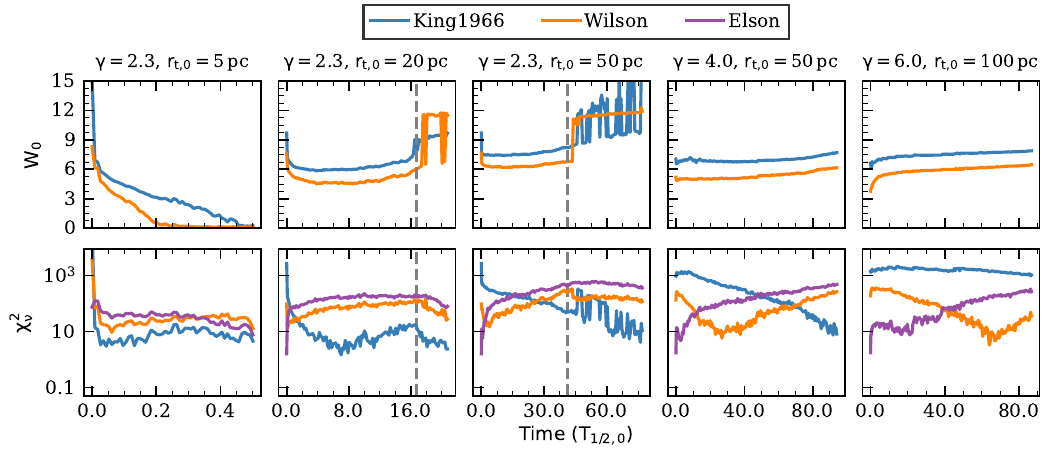}
 \caption{The best-fit values of $\wo$ for King and Wilson profiles and the goodness of fit ($\chisq$) of those profiles (as well as the EFF profile, not shown) for a few representative clusters in \grida. All cluster models shown here have an initial virial radius of $1 \, \pc$. Clusters with tidal radii smaller than $100 \, \pc$ become King-like before the simulation terminates. Conversely, clusters with $\rt = 100 \, \pc$ have not had sufficient time to transition to a King profile and are best-fit by Wilson profiles.  A vertical grey dashed line indicates when the cluster undergoes core collapse (i.e. $\nbh < 5$), if applicable, and a distinct rise in $\wo$ follows this event. A pre-core collapse $\wo$ (value in bracket) is recorded in Table~\ref{tab:transitions} for clusters that undergo core collapse.}
 \label{fig:chisq}
\end{figure*}

Star clusters experience the loss of constituent bodies due to stellar evolution-induced mass loss and dynamical evolution \citep{Lamers2010}. Stellar evolution results in the ejection of compact objects if they acquire a sufficient kick velocity at birth. The dynamical evolution of a stellar system can result in the ejection of objects through various mechanisms, such as  strong encounters between two or more bodies, kicks from tidal disruption events (TDEs) or GW-driven mergers, two-body relaxation, and tidal stripping of stars from a cluster embedded in a tidal field  \citep[see section 2 of][and references therein for a discussion of these mechanisms]{Weath2023}. The clusters in \grida\ are simulated up to the current age of the universe (13.8 \gyr). These clusters lose a significant number of objects through the tidal boundary during the evolution, and our highly tidally truncated clusters do not survive until the present age of the universe. We consider a cluster to be tidally disrupted when its mass drops below $15000 \, \msun$ (5\% of its initial mass of $5\times 10^5 \, \msun$). We do not fit density models to a cluster beyond its disruption time ($\tdisrupt$). 

The King and Wilson models are defined by the fit parameters $\wo$ and $r_{\rm 0}$, while the EFF model uses $\gamma$ and $a$. All models include the central surface brightness as a shared fit parameter, yielding three fit parameters per model. During fitting, the contribution of $\chisq$ from bins beyond $10\,r_{\rm hm}$ --- where $r_{\rm hm}$ is the half-mass radius enclosing 50\% of the mass of the cluster --- is down-weighted by a factor of $1/r^2$. For a cluster at each snapshot, we fit all the three models to the surface brightness profile data and record the best-fit $\chisq$ values for each model. Whichever model has the lowest $\chisq$ is considered the best descriptor of the cluster at that particular epoch. We also record the best-fit values of the model parameters for all the three models at these epochs.

The high density central regions of self-gravitating systems, such as star clusters, undergo a rapid and significant contraction. This collapse of their core happens due to the combined effect of two-body relaxation \citep{Heggie2003} and mass segregation \citep{Spitzer1987}. 
Without counteracting mechanisms, massive objects tend to migrate towards the center of the gravitational potential, deepening the potential well and eventually leading to the collapse of the cluster. Binary systems within the core can undergo strong encounters with passing stars, resulting in the transfer of energy. During this process, the interacting star is expelled from the core, absorbing energy from the binary. The binary tightens due to the reduction in its energy, and the energy imparted to the interacting star is ultimately deposited in the outer halo of the cluster. This process of ``binary burning" counteracts the gravitational collapse \citep[e.g.,][]{Gao1991,Wilkinson2003,Chatterjee2010}. BH binaries are generally the most massive systems with high potential energy and serve as efficient heat sources in this process.

However, BHs can be quickly ejected from cluster cores at the time of formation (primarily due to natal kicks) or later in their evolution through strong dynamical encounters \citep[][]{Morscher2015}. If a cluster loses most of its BHs from the core, it loses the ability to remain heated and may eventually collapse anyway. For our analysis, we classify a cluster as ``core collapsed'' if it retains fewer than 5 BHs. Our simulations show that clusters with \( N_{\rm BH} < 5 \) have a core-to-virial radius ratio (\( r_c/r_v \)) of approximately \(\lesssim 0.3\), whereas clusters with \( N_{\rm BH} \geq 5 \) exhibit \( r_c/r_v \gtrsim 0.3 \). This distinction clearly demarcates core-collapsed clusters from those that are uncollapsed. This is in alignment with clusters with fewer than 10 BHs undergoing a core collapse, as suggested in \cite{Kremer2018}.

The core collapse manifests itself as a noticeable shift in the central surface brightness profile from flat core to a steep cusp. Some of the clusters with $\rv = 1 \, \pc$ undergo core collapse, while none of those with $\rv = 4 \, \pc$ experience it. Within 13.8 Gyr, when a cluster undergoes core collapse, the best-fit King and Wilson parameter, $\wo$, increases sharply (see Figure~\ref{fig:chisq}). Previous studies \citep[e.g.,][]{MG2003LMC, Zocchi2012, Bhat2024} have demonstrated that King models may not fully capture the characteristics of core-collapsed clusters, particularly because the innermost 'cuspy' regions of their surface brightness profiles are better described by a power-law model. However, our simulations reveal that among the models examined (King, Wilson, and EFF), the King model provides the most satisfactory overall fit to the analyzed snapshots of core-collapsed clusters. Notably, the power-law approach fails to account for the sharp decline in surface brightness in the cluster's outer regions, especially in areas with extremely low surface brightness that are typically inaccessible in observational data. Given that our analysis involves fitting the complete cluster profile rather than focusing solely on the innermost regions, we find that the King model performs better for core collapsed clusters, despite its known shortcomings in reproducing the inner cusp.

\section{Transition from YMC to GC}\label{sect:proftrans}
A star cluster, which begins its life having an EFF profile, starts immediately losing mass through its tidal boundary (which is not considered in the EFF density profile). Tidal mass loss through the outskirts coupled with the internal dynamical evolution changes the mass distribution of the cluster.  Therefore, for a given cluster, the values of the fit parameters for density models change as the cluster evolves in time. 

\begin{figure}[!t]
 \centering
 \includegraphics[width=\columnwidth]{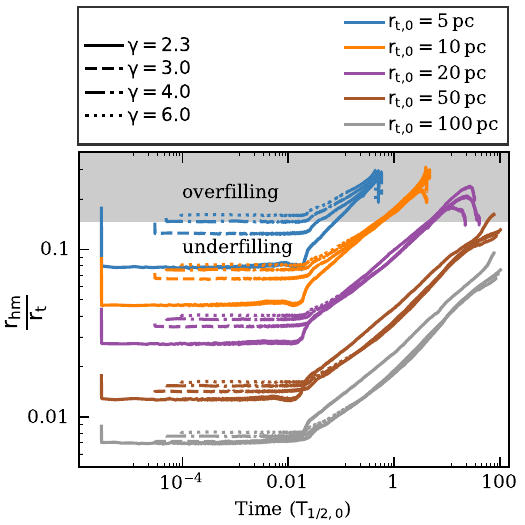}
 \caption{The evolution of the ratio of half-mass radius ($r_{\rm hm}$) and the instantaneous tidal radius ($r_{\rm t}$) of clusters with time. All cluster models shown here have an initial virial radius of $1 \, \pc$. Tidally overfilling clusters ($r_{\rm hm}/ r_{\rm t} > 0.145$) are denoted by the shaded region.  Notably,  clusters with initial tidal radii ($\rt$) smaller than $100 \, \pc$ become King-like before the simulation terminates. Conversely, clusters with $\rt = 100 \, \pc$ have not had sufficient time to transition to a King profile and are best-fit by Wilson models.}
 \label{fig:rhmrt}
\end{figure}

Figure~\ref{fig:chisq} shows the evolution of the best fit parameter, $\wo$, for both King and Wilson models, and $\chisq$ values of the best-fit models for King, Wilson, and EFF models at each snapshot for for a few representative clusters in \grida\  with $\rv = 1 \, \pc$. Typically, a cluster first transitions from an EFF profile to a Wilson profile, and eventually to a King profile. However, highly tidally truncated clusters (e.g., $\rt = 5 \, \pc$) immediately transition to a King profile barely going through a Wilson phase. On the other hand, tidally underfilling clusters (e.g., $\gamma = 6.0,\, \rt = 100$) never transition to a King profile. Table~\ref{tab:transitions} records the disruption times, transition times for EFF $\rightarrow$ Wilson and EFF $\rightarrow$ King, and the King profile parameter $\wo$ of the cluster if it has a King profile at the end of the simulation or its lifetime.

None of our clusters with $\rv = 1 \, \pc$ and tidal radii $\le 20 \, \pc$ survive to the present day (13.8 \gyr), and spend most of their life best fit to a King profile. All clusters with tidal radii $50 \, \pc$  survive and exhibit King profiles at the present day. The survival of these $\rt \geq 50 \, \pc$ clusters to the present day ($\sim 90-100\,\thalf$) approaches the evaporation times of isolated single-mass clusters \citep[c.f. equation 7.141 and its discussion in][]{Binney2008}, where $\thalf$ is the half-mass relaxation time of the cluster at t=0. With the exception of the $\gamma=2.3$ cluster, they spend approximately half of their existence better fit to a Wilson profile than a King profile. None of the clusters with $\rt = 100 \, \pc$ adopt King-like profiles by the present day. Initially, the density of an EFF profile cluster has a constant slope ($\gamma$) at large radii. The dynamic evolution of the cluster begins to flatten the surface brightness profile in the inner parts of the cluster, while tidal stripping in the outskirts brings about a sharp drop in the slope of the outer density profile over, with the density profile near the tidal boundary dropping to zero because the \cite{King1966} profile is derived with an explicit energy cutoff (corresponding to the tidal boundary of the cluster).

\begin{figure*}[!htbp]
 \centering
 \includegraphics[trim={0in 0.0in 0in 0.0in}, clip=True, width=\textwidth]{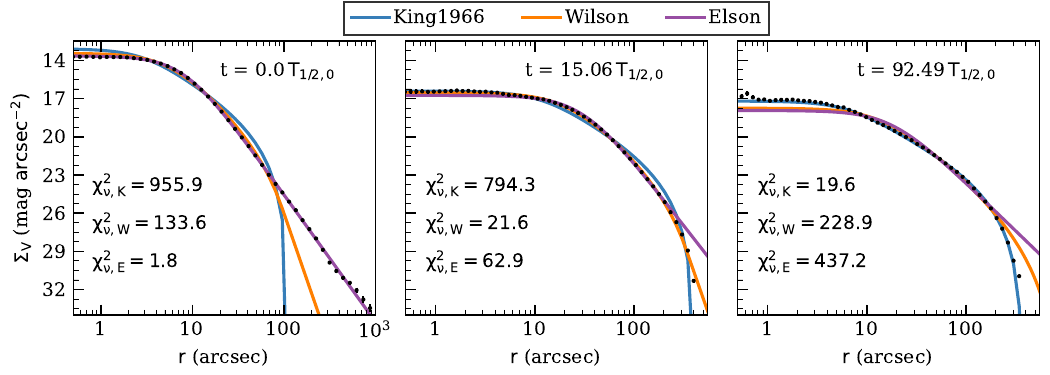}
 \caption{The transition of an EFF profile with time into a King profile for a cluster with $\gamma = 4.0$, $\rt = 50 \, \pc$, and an initial virial radius of $1 \, \pc$. The radial location on the x-axis is given in arcseconds, assuming that the clusters are viewed at a nominal distance of 15 kpc. The \textit{Left panel} shows the surface brightness profiles (open circles) and the corresponding fits for King, Wilson, and EFF models (color lines) at the beginning of the simulation. The \textit{middle panel} and the \textit{right panel} show the same information when the model has evolved to $\sim 15.1 \, \thalf$ and $\sim 92.5 \, \thalf$, respectively. It is visually evident that the EFF, Wilson, and King models fit the surface brightness profile the best in the left, middle, and right panels, respectively.}
 \label{fig3c}
\end{figure*}

The transition of a cluster from an EFF profile to a Wilson or King profile depends on whether the cluster is tidally underfilling or overfilling. A classic definition of tidal overfilling for a single mass model is when $r_{\rm hm}/r_t>0.145$ (following \citealt{Henon1961}) where $r_{\rm hm}$ and $r_{\rm t}$ are the half-mass radius (the radius containing 50\% of the mass of the cluster) and the instantaneous tidal radius of the cluster, respectively. A cluster with a higher $r_{\rm hm}/r_{\rm t}$ ratio is considered tidally overfilling and should lose mass through the tidal boundary. 
Conversely, a cluster with a lower value of $r_{\rm hm}/r_t$ would be underfilling, and potentially not have evolved to a King model by the present day.  As an example, \cite{Ye2022} was best able to reproduce the GC 47 Tuc by starting with an EFF profile with an initial profile slope of $\gamma = 2.1$. Their EFF model best fit the observational data of 47 Tuc at an age of $10.55\,$Gyr exhibiting a ratio of $r_{\rm hm}/r_t = 0.0436$ at that time. According to the above classic definition, this small value reaffirms that 47 Tuc is tidally underfilling and has not transitioned from an EFF to a King profile. 
Figure~\ref{fig:rhmrt} shows the evolution of $r_{\rm hm}/ r_{\rm t}$ for the \grida\  clusters. As a cluster evolves, dynamical heating causes the cluster to expand, increasing the $r_{\rm hm}/r_{\rm t}$ ratio. All clusters with $\rt < 50\,\pc$ become tidally overfilling at some point before the end of the simulation and therefore should transition to a King profile. Indeed, we find that the $\chisq$ value for the fitted King model is smallest for any of the three models around the same time when the cluster enters the overfilling region in the $r_{\rm hm}/r_{\rm t} - \thalf$ plot. In Figure~\ref{fig:rhmrt}, most of the $r_{\rm t} = 50$ clusters never reach the overfilling region. However, based on $\chisq$ values, all $r_{\rm t} = 50$ clusters transition to the King profile before the end of the simulation. This discrepancy is likely due to the use of the classic definition of overfilling for our clusters. Our clusters have a realistic initial mass function (IMF) and undergo stellar evolution resulting in varying masses of cluster constituents. Consequently, the classic definition for overfilling, which assumes single-mass constituents and self-similar cluster models, is not strictly valid for these clusters.  

For $\rv =1\,\pc$ clusters, the transition time among clusters with the same tidal radius increases with a higher slope of the EFF profile. This happens because of two primary reasons $-$ 1) smaller-$\gamma$ clusters are more extended, i.e. they have larger $r_{\rm hm}$ and evaporate comparatively faster in the outer regions.  2) shallow EFF profiles have a higher central density, resulting in shorter relaxation times in the core region \citep[c.f. equation 2.62 of][]{Spitzer1987}. Therefore, clusters with small $\gamma$ undergo a faster transformation in their inner structure as they are dynamically more active. Similarly, for a given slope of the EFF profile, clusters with larger tidal radii experience longer transition times because they evaporate slowly from the outskirts. Figure~\ref{fig3c} shows the evolution of a typical cluster ($\gamma = 4$, $\rt = 50 \, \pc$, $\rv = 1 \, \pc$) surface brightness profile at its three distinct evolutionary phases. This cluster transitions to a Wilson profile at $\sim 13.44 \, \thalf$ and eventually to a King profile at $\sim 70.98 \, \thalf$.  The general trends vis-\`a-vis the tidal radius and the slope of the EFF profile  are similar for clusters with virial radii of $1 \, \pc$ and $4 \, \pc$. For the same $\rt$ and $\gamma$, a cluster with a smaller virial radius has a shorter relaxation time and therefore restructures its central regions faster. Therefore, the $\rv = 4 \, \pc$ clusters have higher transition times for EFF $\rightarrow$ Wilson and EFF $\rightarrow$ King.

\begin{figure*}[!htbp]
 \centering
 \includegraphics[width=\textwidth]{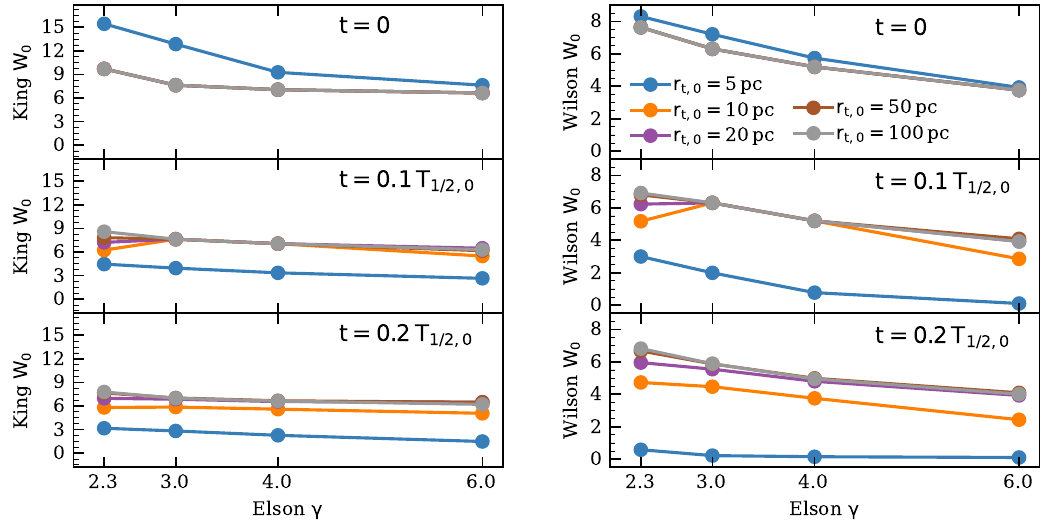}
 \caption{The best-fit $W_0$ parameters for both King (left) and Wilson (right) profiles as a function of initial EFF $\gamma$ parameter.  We show the best-fit $W_0$ parameters at 0, 0.1, and $0.2\,\thalf$ from the cluster initialization.  After $0.2\,\thalf$ the $W_0$ parameter of the cluster is largely dominated by the tidal radius, erasing the initial information about the power-law slope of the initial cluster.}
 \label{fig3d}
\end{figure*}

However, while the general trend in Figure~\ref{fig3c} is that clusters take many half-mass relaxation times to transition to their most ``King-like'' or ``Wilson-like'', it only takes a fraction of a $\thalf$ for a cluster to become more closely modeled by either profile than by it's birth EFF profile.  This is clearly seen in the bottom row of Figure~\ref{fig:chisq}.  In Figure~\ref{fig3d}, we show the best-fit King and Wilson $W_0$ parameters of the clusters as a function of the initial EFF $\gamma$ parameter.  Within 0.2$\,\thalf$, the tidal truncation of the outer parts of the cluster largely dominated the best-fit $W_0$ parameter for both profiles, erasing the initial correlation between $\gamma$ and $W_0$.  This suggests that, after only a fraction of the initial half-mass relaxation time, the $W_0$ parameter is no longer a strong indicator of the initial profile of the cluster (for systems with the same virial radii).  

\section{Runaway collisions and IMBH formation}\label{sect:runaways}
Since collisional runaways primarily occur in the earliest points of cluster evolution, clusters in \gridb\  are evolved for a maximum of 50 Myr. Once a BH forms from the collapse of its stellar counterpart, it has the potential for further growth through collisions with other stars or black holes. However, we are focused on the formation of massive black holes through the collapse of a single massive star formed via stellar runaway collisions. The growth of black holes from mergers of smaller black holes, extensively discussed in other works \citep[e.g.,][]{Freitag2006, Umbreit2012, Carl2018, Carl2019, Fragione2020, Fragione2020_1, Fragione2022}, is not considered here.  At a fixed virial radius, EFF clusters with smaller $\gamma$ have denser and smaller cores (as seen in the right panel of Figure~\ref{fig3a}). Therefore, we expect such clusters to form the most massive IMBHs. Moreover, mass segeregation during cluster evolution enhances the core density, further facilitating stellar encounters.

\begin{figure*}[!ht]
\centering
 \includegraphics[height=0.94\textheight, trim={2.5 2.45in 0in 2.55in}, clip=True]{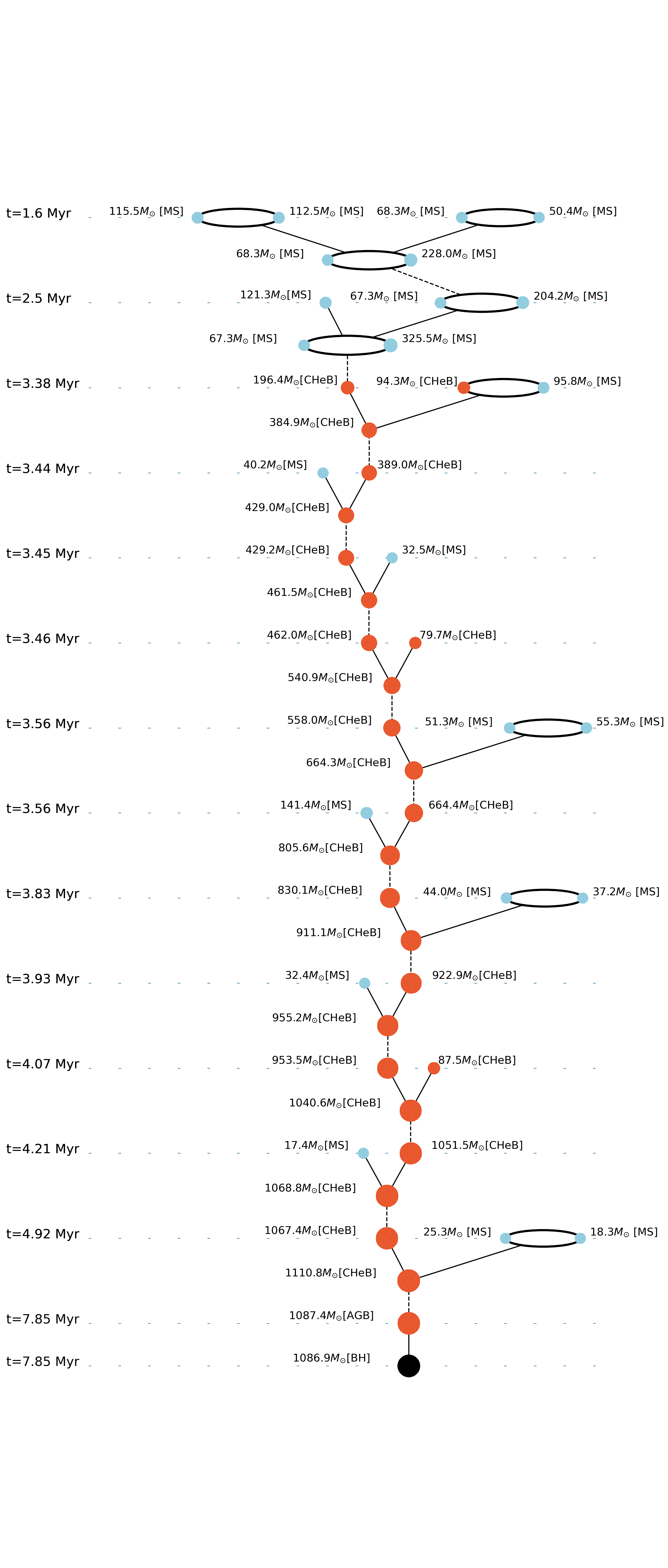}
 \caption{ The collision tree for a $1086.9 \, \msun$ IMBH in a star cluster simulation with $\rv = 2 \, \pc$, Elson $\gamma$ = 2.3, and $\fbh = 1.0$. The size of the markers is proportional to the mass of the star, and binary stars are depicted in elliptical orbits. The MS stars are represented by blue, advanced evolutionary phases such as Core Helium Burning (CHeB) and AGB stars by red markers, and BHs by black markers. The physical time (in \myr) is shown on the left as the runaway collision progresses. The formation of this IMBH involves 1250 single-single, 11 binary-single, and 1 binary-binary encounters. Mergers with stars of mass $< 10 \, \msun$ are not shown in this plot and are cumulatively represented by dashed lines.}
 \label{fig:colltree}
\end{figure*}

\begin{figure*}[!ht]
 \centering
 \includegraphics{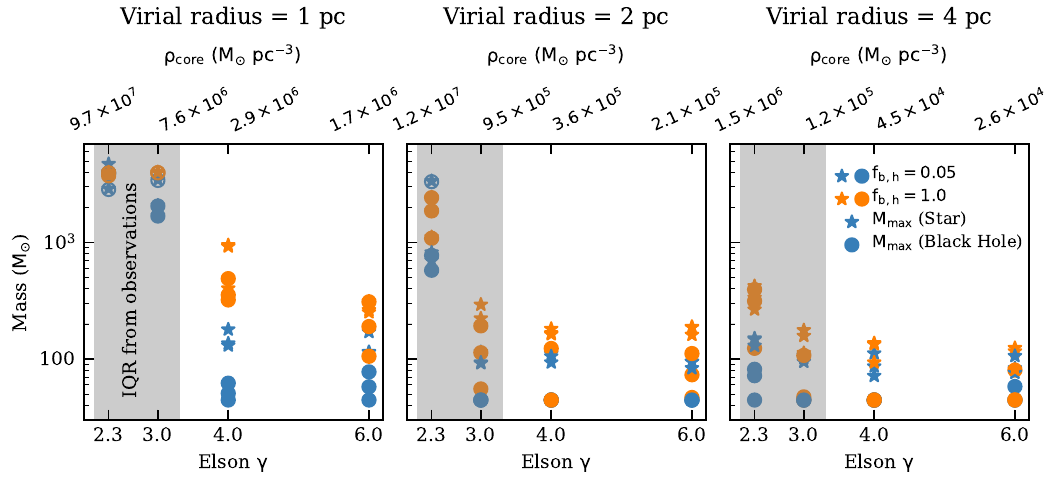}
 \caption{The maximum stellar and BH masses for $\ncluster = 10^6$ star cluster simulations with the CE0 prescription. Each panel corresponds to simulations with different initial virial radii. Within each panel there are six simulations for each $\gamma$, representing three independent realizations for each high-mass binary fraction $\fbh$. The upper horizontal axis displays the core density of the star cluster for a given $\gamma$. The shaded region indicates the Inter Quartile Range (IQR) (range between $25^{\rm th}$ and $75^{\rm th}$ percentile) of $\gamma$ values  of the suggested universal CDF of EFF slopes for young star clusters from \cite{Grudic2021}. Open circles represent cluster simulations where --- 1) either a $4000 \, \msun$ star forms, or 2) the simulation terminates prematurely due to terminal energy error and the massive star has not collapsed by that time. For the \textit{Left panel},  all clusters with $\gamma = 2.3$ either form a $4000 \, \msun$ star or encounter premature termination due to terminal energy error. Note that the most massive star in a simulation does not necessarily collapse to become the most massive BH in that simulation.}
 \label{fig:mstarmbh_ce0}
\end{figure*}

\begin{figure*}[!ht]
\centering
 \includegraphics{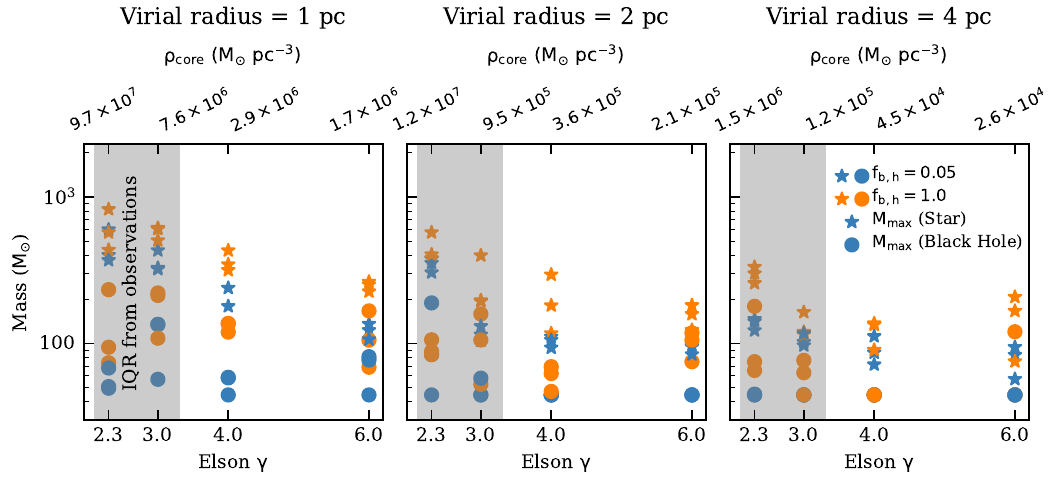}
 \caption{Same as Figure \ref{fig:mstarmbh_ce0}, but for the CE2 prescription. As expected, the masses are consistently smaller for the CE2 prescription compared to the CE0 prescription. However,  the general trend with $\gamma$ and the virial radius remains similar for both prescriptions.}
 \label{fig:mstarmbh_ce2}
\end{figure*}

The H\'enon monte Carlo method becomes unreliable when the mass of a single object within the cluster becomes a significant fraction of the total mass of the cluster. Given our clusters' initial masses of $3\times 10^5$ and $6\times 10^5 \, \msun$, with mass loss during evolution, we decided on a maximum cut-off mass of $4000 \, \msun$ 
for a single object in the cluster. Therefore, we interrupt the simulations of clusters that form stars more massive than $4000 \, \msun$. Furthermore, some simulations terminate prematurely when a massive object ($\sim 1000 \, \msun$) sinks very close to the center of the cluster potential ($<0.005 \, \pc$). In such cases, the potential energy of the cluster changes abruptly within a single time step due to the spherical symmetry of the code, causing the code to crash due to a significant jump in the energy.  We indicate these in Tables~\ref{tab:n1e6} and \ref{tab:n5e5} of the Appendix.  In simulations where the most massive star has not collapsed into a BH before premature termination (marked by $*$ and $\dag\dag$ in Tables~\ref{tab:n1e6} and \ref{tab:n5e5}), we subject it to standalone evolution (disregarding dynamical interactions) in $\cosmic$ for a duration of up to $50\,\myr$ by taking its instantaneous properties at the time of simulation termination as the initial conditions.  We reiterate that because our simulations are terminated whenever a star becomes more massive than $4000M_{\odot}$, that our BHs with masses in that range should be considered as lower limits.  

\begin{figure*}[!ht]
\centering
 \includegraphics{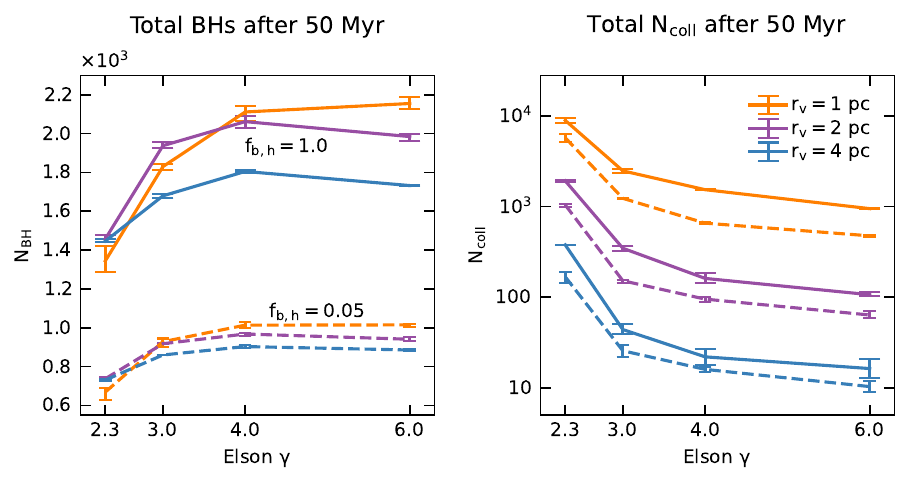}
 \caption{\textit{Left panel:} shows the number of BHs in $\ncluster = 10^6$ cluster simulations for the CE2 prescription. Solid (dashed) curves represent the average number of BHs for each $\gamma$ with $\fbh$ of 1.0 (0.05). Each data point shows the range of $\nbh$ formed in clusters of a given $\gamma$ for three independent realizations. Clusters with higher core densities (smaller $\gamma$), where massive BHs form and grow rapidly, tend to form a smaller number of BHs. \textit{Right panel:} shows the total number of collisions upto $50\, \myr$ for these simulations.}
 \label{fig:nbh_gamma}
\end{figure*}

Figure~\ref{fig:colltree} shows the merger tree for a typical IMBH formed in our simulations. The interaction of two binaries with MS stars and component masses of ($M_{\rm 1}, M_{\rm 2}$) = ($115.5 \, \msun$, $112.5 \, \msun$) and ($68.3 \, \msun$, $50.4 \, \msun$) marks the beginning of the runaway process. This interaction results in the coalescence of the first binary into a $228 \, \msun$ MS star, which captures the $68.3 \, \msun$ star from the second binary. The product binary then collides with a MS single star of mass $121.3 \, \msun$ and the primary star merges with the incoming single star, forming a $325.5 \, \msun$ primary star. A series of further single-single and binary-single collisions lead to the formation of a $1087.4 \, \msun$ asymptotic giant branch (AGB) star, which then collapses into an IMBH of mass $1086.9 \, \msun$. The formation of this IMBH involves 1250 single-single, 11 binary-single, and 1 binary-binary encounters. Single-single, binary-single, and binary-binary encounters contribute 54.4\%, 37.2\%, and 8.4\% of the total mass growth through collisions leading up to the formation of the IMBH, respectively.

We tabulate key simulation parameters and BH formation results in Table~\ref{tab:n1e6} (Table~\ref{tab:n5e5}) in the Appendix for our collisional runaway grid for clusters with $10^6$ ($5\times10^5$) stars. We used two different prescriptions for collisions involving giant stars, CE0 and CE2, discussed in Section~\ref{sect:sims}. In the CE2 prescription, more envelope mass is lost during a collision involving giant stars compared to the CE0 prescription. Therefore we see that stars in the CE0 prescription simulations are generally more massive than the CE2 prescription simulations and form the most massive IMBHs. In Figures~\ref{fig:mstarmbh_ce0} and \ref{fig:mstarmbh_ce2}, we display the masses of the most massive stars and BHs formed in our simulations with $10^6$ particles for the CE0 and CE2 prescription, respectively.

For a given cluster size and profile slope $\gamma$, clusters with higher initial binary fraction ($\fbh$) form more massive IMBHs due to increased binary-mediated stellar collisions, aligning with the findings of \cite{Gonzalez2021}. For a given cluster size and initial $\fbh$, clusters with denser cores (smaller $\gamma$) form more massive IMBHs. Conversely, clusters with larger initial virial radii have fewer interactions for objects in the core and form less massive (or no) IMBHs. The most massive stars formed in the CE0 prescription simulations with $N = 10^6$  are $4000 \, \msun$ (our simulation termination criterion) and the most massive BHs have similar masses. For the CE2 prescription simulations, the most massive star has a mass of $824 \, \msun$ which collapses into a $232.8 \, \msun$ IMBH. 

All simulations in \gridb\  form several BHs in the upper mass gap within any given cluster. Multiple IMBHs are observed in 13 out of 72 clusters for $\ncluster=10^6$ ($\mcluster \simeq 6\times10^5 \, \msun$) clusters using the CE0 prescription, with the highest number (7) occurring in the cluster with $\gamma = 4.0$, $\rv = 1 \, \pc$, $\fbh = 1.0$. Whereas, for $\ncluster=10^6$ and the CE2 prescription simulations, only 1 out of 72 clusters forms more than one IMBH. For $\ncluster=5\times10^5$, the number of clusters forming multiple IMBHs within a single cluster is 5 and 1 for the CE0 and CE2 prescriptions, respectively. Notably, the highest number of IMBHs forms in a $\gamma = 4.0$ cluster, while the most massive IMBHs form in $\gamma = 2.3$ clusters.

\subsection{Consequences for Long-term Cluster Evolution}
From Tables~\ref{tab:n1e6} and \ref{tab:n5e5}, it becomes apparent that clusters with smaller values of $\gamma$, while keeping other initial conditions fixed, produce considerably fewer stellar-mass BHs. 
For instance, our $\ncluster=10^6$, $\fbh = 0.05$, $\rv=1\,\pc$ clusters with $\gamma=2.3$ form $\sim20\%$ fewer stellar-mass BHs compared to $\gamma=6.0$ clusters. Dynamical interactions among these BHs often result in their expulsion from the cluster core \citep[e.g.,][]{Breen2013, Morscher2015}. 
The right panel of Figure~\ref{fig:nbh_gamma}, shows the cumulative number of collisions ($N_{\rm coll}$) in $\ncluster = 10^6$ star simulations with the CE2 prescription at $50\,\myr$. For $\fbh = 0.05$, the ratio of average $N_{\rm coll}$ between $\gamma = 2.3$ and $\gamma=6.0$ clusters is 12.06, 16.37, and 16.26 for $\rv=1$, 2, and $4\,\pc$ clusters, respectively. For $\fbh = 1.0$, this ratio is 9.4, 18.02, and 23.14, respectively. These collisions are largely between massive stars, the progenitors of stellar-mass BHs, meaning that smaller-$\gamma$ clusters produce fewer BHs. This is evident from the left panel of Figure~\ref{fig:nbh_gamma} which shows the total number of BHs retained at $50\, \myr$ in these clusters. For $\fbh = 0.05$, the ratio of average $\nbh$ between $\gamma = 2.3$ and $\gamma=6.0$ clusters is 0.66, 0.78, and 0.82 for $\rv=1$, 2, and $4\,\pc$ clusters, respectively. For $\fbh = 1.0$, this ratio is 0.62, 0.73, and 0.83, respectively. The total number of BHs remaining in $\fbh$ = 1.0 simulations is consistently higher than in $\fbh$ = 0.05 simulations.  It is interesting to note that $\gamma \leq 3.0$ star clusters form and retain fewer stellar-mass BHs compared to $\gamma=6.0$ clusters of the same size but the most massive IMBHs are formed in these centrally dense star clusters. Although not shown, similar trends are exhibited by CE0 simulations.

As has been shown previously \citep{Breen2013,Kremer2018,Kremer2020a}, the eventual \emph{observational} core collapse of clusters is largely driven by the expulsion of their stellar-mass BH populations.  This can occur in clusters that either dynamically eject their BHs by the present day \citep[if they have sufficiently short $\thalf$, e.g.][]{Kremer2019}, or in clusters that are born with fewer BHs to begin with \citep[e.g.,][]{Rodriguez2023}.  What this suggests is that the very process of creating an IMBH through runaway collisions can deplete the cluster of stellar-mass BHs, and significantly accelerate core collapse!  Ironically, our results are in agreement with the conventional wisdom that core-collapsed clusters may host IMBHs due to their extreme central densities.  However, instead of the central density being responsible for IMBH formation, we argue that it is IMBH formation that leads to core collapse, and the high central density!  Of course, just because a cluster lost its BH-progenitor stars in a collisional runaway does not mean that the cluster still contains an IMBH today \citep[see][which argues that IMBHs born with masses in the range $120-500\,\msun$ are ejected from the clusters within the first $\sim 500\,\myr$ of evolution]{Gonzalez2022}.

Our findings also align with previous studies by \cite{Hong2020} and \cite{Rujuta2024}, which demonstrate that the rate of binary BH mergers in clusters decreases with higher central densities and stronger runaway processes. This is in contrast with the slow scenario of IMBH formation, where the initial core collapse of the cluster increases the central density \citep[e.g.,][]{Greene2020, Gieles2019}, facilitating a higher number of collisions and the eventual formation of an IMBH. In our fast scenario, the presence of an IMBH expedites the core collapse process within the cluster.

\section{Discussion and Conclusions}\label{sect:discussion}

In this work, we have studied the dynamical evolution of young star clusters which are born with an EFF density profiles, similar to observations of young star clusters in the local universe. We have used two simulation grids with diverse sets of cluster initial conditions - \grida\  consisting 40 star cluster models to study the evolution of young EFF profile to old King profiles, and  \gridb\  consisting 288 star cluster models to explore the formation of IMBHs. In particular, we focus solely on the evolution of the cluster and formation of massive BHs within the initial 50 \myr.  Our main conclusions are:

\begin{enumerate}[label={\arabic*)}, leftmargin=*]
    \item We have demonstrated that initially tidally overfilling EFF clusters naturally transform to  King or Wilson profiles through dynamical evolution. The transition from an EFF to a Wilson or King profile depends on factors like tidal filling, virial radius, and the initial slope of the EFF profile. While it may take many half-mass relaxation times for a cluster to fully conform to a King or Wilson profile, it loses its initial EFF profile characteristics much sooner. This suggests that the King / Wilson parameter $\wo$ of most clusters at the present day are not a good indicator of a cluster's initial profile or potential for IMBH formation. To our knowledge, this is the first study to demonstrate the evolution from EFF to Wilson or King profiles in self-consistent numerical models of star clusters using state-of-the-art dynamics and stellar evolution prescriptions.   

    \item We find that massive EFF clusters with shallow density profiles can produce runaway stellar mergers with masses as high as $824\,\msun$ ($4000\,\msun$) for CE2 (CE0) prescription for the treatment of giant star envelopes during collisions. These massive stars collapse to form IMBHs with masses an order of magnitude higher than IMBHs produced in recent studies which assumed a King profile \citep{Gonzalez2022, Gonzalez2021, Kremer2020}. This result is significant because it demonstrates that using EFF profiles, which better represent observational data of YMCs, leads to more massive IMBHs. We emphasize that the CE0 prescription serves as an upper limit on the IMBH masses produced by runaway processes in our simulations. Meanwhile, because the CE2 prescription treats every giant star collision as involving a common-envelope phase, it may underestimate the final mass in cases where a full common envelope does not develop. Consequently, the true outcome likely lies between the masses predicted by the CE0 and CE2 prescriptions.
    
    Additionally, we observe multiple clusters hosting several IMBHs, with the number of such clusters showing a strong correlation with the initial mass of the cluster, the high-mass binary fraction, and the collision prescription utilized for interactions involving giant stars. This is in-line with previous studies.
    
    \item In centrally dense clusters characterized by smaller values of $\gamma$, the formation and growth of a collisional runaway star (and a subsequent IMBH) can reduce the number of stellar-mass BHs by nearly 40\%.  This trend suggests that the formation and retention of fewer stellar-mass BHs in clusters with smaller $\gamma$ values may accelerate the process of core collapse within these clusters. This in turn suggests that core-collapsed clusters may have been the site of IMBH formation at some point.
    Our results align with the conventional wisdom that core-collapsed clusters host IMBHs due to their high central densities. However, instead of the core-collapse being responsible for IMBH formation, we argue that it is IMBH formation that leads to earlier core collapse.

\end{enumerate}

Our study has relied on several assumptions for the outcomes of stellar collisions and massive star evolution, which we must address.  First, we adopted two different collision prescriptions for stellar collisions involving giant stars and, unsurprisingly, find that more massive black holes are formed when using the ``sticky sphere" collision prescription; i.e., CE0. However, the reality likely lies somewhere between the two collision prescriptions, CE0 and CE2. Our work emphasizes the importance of understanding the collision mechanism of massive stars. Accurate modeling of giant stellar collisions is highly uncertain, but recently it has been attempted by \cite{Costa2022} and \cite{Ballone2022} using hydrodynamical codes. However, these studies are limited in scope as they focus on specific encounter scenarios. Further work is required to accurately classify the properties of merger products in general collision scenarios, involving various stellar types, stellar masses, impact parameters, and collision velocities.   

When a star comes in close proximity to a compact object, the outcome of the interaction depends on the impact parameter, as well as the masses, types, and approach velocities of the interacting objects. \cmc\ does not yet have an extensive formalism to accurately model such collisions. By default, a star colliding with a BH in \cmc\ results in the complete disruption of the star, termed a TDE \citep[see][for a review of TDEs]{Gezari2021}. However, this is not a realistic prescription if, for instance, a $5 \, \msun$ BH with a high relative velocity shoots through the loosely bound envelope of a giant star, as the giant star will probably lose only a small fraction of its mass. Therefore, we have used an ad hoc prescription (described in section~\ref{sect:sims}) to avoid some of these situations. However, we emphasize that accurate modeling of TDEs is necessary, and efforts are currently underway.

Massive stars can undergo thousands of collisions, resulting in the formation of massive stars that eventually collapse to form an IMBH. When two stars collide, the chemical composition of the resulting merger product changes as a result of the mixing of material from the stars. Consequently, the merged product exhibits a chemical composition distinct from that of either progenitor. The extent of mixing determines the enhancement in the lifetime of a star \citep{Hurley2002, Breivik2020}. A prolonged lifetime for a massive star would allow for a greater number of collisions with other objects, thereby increasing the mass of the remnant formed during the star's eventual collapse.     

We form stars with masses up to $4000 \, \msun$ in our simulations, employing the stellar evolution prescriptions of \cosmic\ \citep{Breivik2020}. These prescriptions are based on interpolation and extrapolation formulae from the \texttt{SSE} and \texttt{BSE} packages, derived by \cite{Hurley2000, Hurley2002}, which, in turn, are based on the grid of stellar evolution tracks computed by \cite{POLS1998} for masses ranging from 0.5 to $50 \, \msun$. Notably, detailed stellar evolution models exist only for stars with a maximum mass of a few hundred solar masses (e.g., BoOST \citep{Szecsi2022}, BPASS \citep{Eldridge2017}, MIST \citep{Choi2016}). The maximum stellar radius during a star's evolution can vary by $1000 \, \rsun$ and the mass of the remnant formed at the end of a star's life can differ by $20 \, \msun$ among these stellar evolution models \citep[Figures 2 and 3 of][]{Poojan2022}. The radius at the main-sequence turnoff, \( R_{\rm TMS} \), is an important indicator of the evolution of a star's physical size during its main-sequence lifetime and its progression into the giant phase. Unfortunately, for very high masses (\( M > 200\,\msun \)) and low metallicities (\( Z \sim 0.01\,\zsun \)), the SSE-based extrapolation can yield nonphysical (i.e., negative) values for \( R_{\rm TMS} \). Nevertheless, for the metallicity employed in our simulations (\( Z = 0.1\,\zsun \)), the \( R_{\rm TMS} \) obtained from SSE-based formulae remains broadly consistent with that from more comprehensive stellar evolution codes, with the variability among these codes being comparable to the difference between SSE and the comprehensive models. Consequently, while interpolation and extrapolation in SSE/BSE provide a practical framework for covering a wide range of masses and metallicities, further refinement is necessary for accurately modeling extremely massive stars at the lowest metallicities. This is critical because accurate modeling of stellar evolutionary processes in massive stars significantly influences both the physical size of stars (which strongly affects collision probabilities) and the mass of the resulting compact remnants. 

Observations of star-forming regions suggest that both the distribution of molecular gas collapsing to form stars and the newly formed stars are asymmetric and clumpy in nature \citep[][and references therein]{Goodwin2004}. Therefore, ideally, the young star clusters should be modeled with fractal initial conditions \citep{Kupper2011}. Using direct N-body simulations of $10^4$ star clusters with masses in the range of $10^3 - 3\times 10^4 \, \msun$  and fractal initial conditions, \citet{Ugo2021}  concluded that runaway stellar collisions efficiently produce BHs with masses $> 100 \, \msun$  and these BHs are efficient in acquiring a companion BH. They found that more massive and metal-poor clusters are more efficient at producing IMBHs. However, using fractal initial conditions in a Hénon-type Monte Carlo code, such as \cmc, which assumes spherical symmetry, is not feasible. Therefore, we have to use a spherically symmetric density profile representative of young star clusters, which may limit the number of collisions that occur naturally in fractal initial conditions.

In this study, we exclusively focused on clusters with 10\% solar metallicity. The metallicity of a star impacts wind mass loss rates, especially for massive stars. Stars with lower metallicity exhibit smaller wind mass loss rates and can retain a larger fraction of their mass. For young star clusters, \cite{Ugo2020} demonstrates that 6\% of BHs in lower metallicity clusters ($Z = 0.0002$) have $M>60 \, \msun$ compared to less than $1\%$ in higher metallicity clusters ($Z = 0.02$). Furthermore, \cite{Shrivastava2022} found that the number of BHs with $M>44.5 \, \msun$ formed in low-metallicity clusters ($Z = 0.1 \, \zsun$) can be 10 times as many as in high-metallicity clusters ($Z = \zsun$).  In future work, we intend to explore the impact of metallicity on the formation and properties of IMBHs by incorporating data on stellar physical sizes from the latest stellar evolution models for massive stars.

\section*{Acknowledgements}
We are grateful to Elena Gonz{\'a}lez and Kyle Kremer for useful discussions. This work was supported by NSF Grant AST-2310362 and NASA ATP Grant 80NSSC22K0722.  CR also acknowledges support from a Charles E.~Kaufman Foundation New Investigator Research Grant, an Alfred P.~Sloan Research Fellowship, and a David and Lucile Packard Foundation Fellowship.


\vspace{0.5 cm}
\appendix
\vspace{-3 em}
\startlongtable
\begin{deluxetable*}{@{\extracolsep{\fill}}lccccccc|lccccccr@{}}
\tablecaption{Relevant simulation parameters and outputs for \gridb\  cluster simulations with $\ncluster = 10^6$ objects.}\label{tab:n1e6}

\tabletypesize{\small}
\tablehead{
 \colhead{$\gamma$} & \colhead{$\rv$} & \colhead{$\fbh$} &
 \colhead{$\nbh$} & \colhead{$N_{\rm PI}$} & \colhead{$N_{\rm IM}$} & 
 \colhead{$M_{\rm BH, max}$} & \colhead{$M_{\rm *, max}$} &
 \colhead{ $\gamma$} & \colhead{$\rv$} & \colhead{$\fbh$} &
 \colhead{$\nbh$} & \colhead{$N_{\rm PI}$} & \colhead{$N_{\rm IM}$} & 
 \colhead{$M_{\rm BH, max}$} & \colhead{$M_{\rm *, max}$}\\[-1ex]
 \colhead{} & \colhead{($\pc$)} & \colhead{} &
 \colhead{} & \colhead{} & \colhead{} & 
 \colhead{($\msun$)} & \colhead{($\msun$)} & 
 \colhead{} & \colhead{($\pc$)} & \colhead{} &
 \colhead{} & \colhead{} & \colhead{} & 
 \colhead{($\msun$)} & \colhead{($\msun$)}\\
\hline
\multicolumn{8}{c|}{\rule[10pt]{0pt}{0pt} \small{CE2}} & \multicolumn{8}{c}{\rule[15pt]{0pt}{0pt} \small{CE0}}
 }
\startdata
& & & & & & & & & & & & &\\[-3ex] 
2.3 & 1 & 0.05 & 2013 & 2  & 0 & 50.7  & 400.5 &     2.3 & 1 & 0.05 & 322  & 0  & 1 & $3965.3^*$ & $4007.5^*$\\
2.3 & 1 & 0.05 & 2009 & 1  & 0 & 49.4  & 598.9 &     2.3 & 1 & 0.05 & 375  & 0  & 1 & $3903.7^*$ & $4719.1^*$\\
2.3 & 1 & 0.05 & 2060 & 2  & 0 & 67.9  & 370.0 &     2.3 & 1 & 0.05 & 817  & 0  & 2 & $2851.2^{\dag\dag}$ & $2895.2^{\dag\dag}$\\
2.3 & 1 & 1.0  & 2821 & 9  & 0 & 73.7  & 572.1 &     2.3 & 1 & 1.0  & 498  & 0  & 1 & $3961.9^*$ & $4000.6^*$\\
2.3 & 1 & 1.0  & 2931 & 9  & 1 & 232.8 & 824.1 &     2.3 & 1 & 1.0  & 356  & 1  & 1 & $3956.5^*$ & $4000.0^*$\\
2.3 & 1 & 1.0  & 2889 & 7  & 0 & 94.2  & 435.2 &     2.3 & 1 & 1.0  & 928  & 0  & 1 & $3740.4^{\dag\dag}$ & $3784.0^{\dag\dag}$\\
3.0 & 1 & 0.05 & 2485 & 2  & 1 & 134.8 & 321.6 &     3.0 & 1 & 0.05 & 1776 & 0  & 1 & 2056.4 & 2074.6\\
3.0 & 1 & 0.05 & 2470 & 0  & 1 & 134.7 & 432.0 &     3.0 & 1 & 0.05 & 2446 & 1  & 1 & 1685.8 & 1691.0\\
3.0 & 1 & 0.05 & 2461 & 4  & 0 & 56.9  & 327.7 &     3.0 & 1 & 0.05 & 461  & 2  & 2 & $3410.2^{\dag\dag}$ & $3453.7^{\dag\dag}$\\
3.0 & 1 & 1.0  & 3713 & 11 & 1 & 212.7 & 614.2 &     3.0 & 1 & 1.0  & 1405 & 5  & 4 & $3957.3^*$ & $4000.1^*$\\
3.0 & 1 & 1.0  & 3635 & 10 & 0 & 108.6 & 505.0 &     3.0 & 1 & 1.0  & 504  & 1  & 3 & $3969.0^*$ & $4002.7^*$\\
3.0 & 1 & 1.0  & 3692 & 10 & 2 & 221.1 & 604.0 &     3.0 & 1 & 1.0  & 421  & 0  & 4 & $3958.6^*$ & $4000.4^*$\\
4.0 & 1 & 0.05 & 2618 & 0  & 0 & 44.5  & 130.8 &     4.0 & 1 & 0.05 & 2627 & 1  & 0 & 50.7 & 130.8\\
4.0 & 1 & 0.05 & 2625 & 1  & 0 & 58.2  & 179.3 &     4.0 & 1 & 0.05 & 2630 & 0  & 0 & 44.5 & 179.3\\
4.0 & 1 & 0.05 & 2621 & 1  & 0 & 58.7  & 239.2 &     4.0 & 1 & 0.05 & 2630 & 2  & 0 & 62.1 & 136.1\\
4.0 & 1 & 1.0  & 4153 & 9  & 1 & 136.4 & 431.5 &     4.0 & 1 & 1.0  & 4093 & 25 & 6 & 354.8 & 403.5\\
4.0 & 1 & 1.0  & 4208 & 6  & 1 & 136.9 & 316.5 &     4.0 & 1 & 1.0  & 4128 & 8  & 5 & 321.4 & 949.1\\
4.0 & 1 & 1.0  & 4088 & 12 & 0 & 119.4 & 346.2 &     4.0 & 1 & 1.0  & 4071 & 13 & 7 & 490.0 & 925.9\\
6.0 & 1 & 0.05 & 2550 & 0  & 0 & 44.5  & 123.0 &     6.0 & 1 & 0.05 & 2545 & 0  & 0 & 44.5 & 170.8\\
6.0 & 1 & 0.05 & 2543 & 2  & 0 & 81.2  & 135.8 &     6.0 & 1 & 0.05 & 2554 & 2  & 0 & 77.6 & 114.3\\
6.0 & 1 & 0.05 & 2554 & 2  & 0 & 77.0  & 106.6 &     6.0 & 1 & 0.05 & 2555 & 1  & 0 & 57.8 & 114.8\\
6.0 & 1 & 1.0  & 4100 & 5  & 0 & 105.3 & 263.4 &     6.0 & 1 & 1.0  & 4158 & 9  & 2 & 190.1 & 263.4\\
6.0 & 1 & 1.0  & 4096 & 2  & 0 & 68.8  & 226.2 &     6.0 & 1 & 1.0  & 4111 & 2  & 2 & 309.8 & 311.8\\
6.0 & 1 & 1.0  & 4181 & 6  & 1 & 166.8 & 252.0 &     6.0 & 1 & 1.0  & 4107 & 9  & 0 & 105.7 & 252.0\\[1ex]
\hline
 & & & & & & & & & & & & &\\[-3ex] 
2.3 & 2 & 0.05 & 2198 & 2   & 0 & 83.9  & 377.3 &     2.3 & 2 & 0.05 & 729  & 1  & 1 & $3324.8^{\dag\dag}$ & $3368.4^{\dag\dag}$\\
2.3 & 2 & 0.05 & 2194 & 2   & 1 & 189.1 & 304.8 &     2.3 & 2 & 0.05 & 2226 & 0  & 1 & 577.9 & 702.0\\
2.3 & 2 & 0.05 & 2181 & 0   & 0 & 44.5  & 354.0 &     2.3 & 2 & 0.05 & 2220 & 0  & 1 & 775.4 & 819.8\\
2.3 & 2 & 1.0  & 3099 & 13  & 0 & 88.3  & 381.3 &     2.3 & 2 & 1.0  & 2558 & 1  & 1 & $2423.2^{\dag}$ & $2433.6^{\dag}$\\
2.3 & 2 & 1.0  & 3127 & 9   & 0 & 105.8 & 572.3 &     2.3 & 2 & 1.0  & 3217 & 3  & 3 & 1086.9 & 1112.6\\
2.3 & 2 & 1.0  & 3097 & 13  & 0 & 83.8  & 404.7 &     2.3 & 2 & 1.0  & 3259 & 2  & 3 & 1866.2 & 1872.3\\
3.0 & 2 & 0.05 & 2489 & 0   & 0 & 44.5  & 104.4 &     3.0 & 2 & 0.05 & 2493 & 0  & 0 & 44.5  & 92.1\\
3.0 & 2 & 0.05 & 2489 & 0   & 0 & 44.5  & 114.8 &     3.0 & 2 & 0.05 & 2494 & 0  & 0 & 44.5  & 114.8\\
3.0 & 2 & 0.05 & 2494 & 1   & 0 & 57.9  & 130.9 &     3.0 & 2 & 0.05 & 2496 & 0  & 0 & 44.5  & 93.3\\
3.0 & 2 & 1.0  & 3871 & 0   & 1 & 158.8 & 398.7 &     3.0 & 2 & 1.0  & 3846 & 4  & 0 & 55.4  & 194.8\\
3.0 & 2 & 1.0  & 3748 & 2   & 0 & 52.6  & 196.7 &     3.0 & 2 & 1.0  & 3852 & 3  & 0 & 113.4 & 223.0\\
3.0 & 2 & 1.0  & 3781 & 6   & 0 & 105.5 & 193.8 &     3.0 & 2 & 1.0  & 3863 & 5  & 1 & 192.7 & 293.2\\
4.0 & 2 & 0.05 & 2627 & 0   & 0 & 44.5  & 110.9 &     4.0 & 2 & 0.05 & 2629 & 0  & 0 & 44.5  & 110.9\\
4.0 & 2 & 0.05 & 2639 & 0   & 0 & 44.5  & 93.4 &      4.0 & 2 & 0.05 & 2633 & 0  & 0 & 44.5  & 93.5\\
4.0 & 2 & 0.05 & 2631 & 0   & 0 & 44.5  & 105.5 &     4.0 & 2 & 0.05 & 2631 & 0  & 0 & 44.5  & 105.5\\
4.0 & 2 & 1.0  & 3755 & 1   & 0 & 69.4  & 117.1 &     4.0 & 2 & 1.0  & 3921 & 2  & 1 & 123.2 & 166.4\\
4.0 & 2 & 1.0  & 3761 & 2   & 0 & 62.3  & 295.0 &     4.0 & 2 & 1.0  & 3744 & 0  & 0 & 44.5  & 163.3\\
4.0 & 2 & 1.0  & 3788 & 4   & 0 & 47.0  & 181.6 &     4.0 & 2 & 1.0  & 3936 & 0  & 0 & 44.5  & 181.6\\
6.0 & 2 & 0.05 & 2539 & 0   & 0 & 44.5  & 94.7 &      6.0 & 2 & 0.05 & 2539 & 0  & 0 & 44.5  & 94.7\\
6.0 & 2 & 0.05 & 2545 & 0   & 0 & 44.5  & 97.5 &      6.0 & 2 & 0.05 & 2538 & 0  & 0 & 44.5  & 95.0\\
6.0 & 2 & 0.05 & 2545 & 0   & 0 & 44.5  & 84.1 &      6.0 & 2 & 0.05 & 2542 & 0  & 0 & 44.5  & 83.5\\
6.0 & 2 & 1.0  & 3507 & 2   & 0 & 117.9 & 182.0 &     6.0 & 2 & 1.0  & 3482 & 2  & 0 & 73.3  & 187.3\\
6.0 & 2 & 1.0  & 3496 & 1   & 0 & 75.1  & 158.5 &     6.0 & 2 & 1.0  & 3454 & 1  & 0 & 46.8  & 161.4\\
6.0 & 2 & 1.0  & 3525 & 1   & 0 & 105.3 & 123.4 &     6.0 & 2 & 1.0  & 3545 & 1  & 0 & 111.5 & 188.9\\[1ex]
\hline
 & & & & & & & & & & & & &\\[-3ex] 
2.3 & 4 & 0.05 & 2138 & 0 & 0 & 44.5  & 146.2 &     2.3 & 4 & 0.05 & 2139 & 1 & 0 & 72.0  & 150.4\\
2.3 & 4 & 0.05 & 2151 & 1 & 0 & 45.1  & 141.1 &     2.3 & 4 & 0.05 & 2146 & 0 & 0 & 44.5  & 147.7\\
2.3 & 4 & 0.05 & 2141 & 0 & 0 & 44.5  & 122.8 &     2.3 & 4 & 0.05 & 2143 & 2 & 0 & 81.7  & 130.1\\
2.3 & 4 & 1.0  & 3008 & 3 & 0 & 75.1  & 332.7 &     2.3 & 4 & 1.0  & 3026 & 4 & 1 & 123.6 & 265.8\\
2.3 & 4 & 1.0  & 3085 & 5 & 1 & 179.2 & 257.7 &     2.3 & 4 & 1.0  & 3046 & 6 & 3 & 314.2 & 336.1\\
2.3 & 4 & 1.0  & 3017 & 2 & 0 & 65.4  & 300.0 &     2.3 & 4 & 1.0  & 3052 & 3 & 1 & 394.4 & 422.4\\
3.0 & 4 & 0.05 & 2451 & 0 & 0 & 44.5  & 96.9 &      3.0 & 4 & 0.05 & 2446 & 0 & 0 & 44.5  & 107.1\\
3.0 & 4 & 0.05 & 2448 & 0 & 0 & 44.5  & 102.1 &     3.0 & 4 & 0.05 & 2441 & 0 & 0 & 44.5  & 102.1\\
3.0 & 4 & 0.05 & 2444 & 0 & 0 & 44.5  & 116.9 &     3.0 & 4 & 0.05 & 2444 & 0 & 0 & 44.5  & 94.8\\
3.0 & 4 & 1.0  & 3065 & 1 & 0 & 76.7  & 119.8 &     3.0 & 4 & 1.0  & 3063 & 0 & 0 & 44.5  & 177.0\\
3.0 & 4 & 1.0  & 3052 & 2 & 0 & 63.1  & 115.0 &     3.0 & 4 & 1.0  & 3104 & 1 & 0 & 47.4  & 112.6\\
3.0 & 4 & 1.0  & 3047 & 0 & 0 & 44.5  & 163.5 &     3.0 & 4 & 1.0  & 3125 & 2 & 0 & 107.5 & 158.2\\
4.0 & 4 & 0.05 & 2608 & 0 & 0 & 44.5  & 71.7 &      4.0 & 4 & 0.05 & 2612 & 0 & 0 & 44.5  & 71.7\\
4.0 & 4 & 0.05 & 2608 & 0 & 0 & 44.5  & 111.9 &     4.0 & 4 & 0.05 & 2602 & 0 & 0 & 44.5  & 111.9\\
4.0 & 4 & 0.05 & 2615 & 0 & 0 & 44.5  & 86.6 &      4.0 & 4 & 0.05 & 2620 & 0 & 0 & 44.5  & 86.5\\
4.0 & 4 & 1.0  & 3307 & 0 & 0 & 44.5  & 136.7 &     4.0 & 4 & 1.0  & 3324 & 0 & 0 & 44.5  & 136.7\\
4.0 & 4 & 1.0  & 3333 & 0 & 0 & 44.5  & 134.2 &     4.0 & 4 & 1.0  & 3334 & 0 & 0 & 44.5  & 134.2\\
4.0 & 4 & 1.0  & 3293 & 0 & 0 & 44.5  & 90.2 &      4.0 & 4 & 1.0  & 3337 & 0 & 0 & 44.5  & 93.5\\
6.0 & 4 & 0.05 & 2520 & 0 & 0 & 44.5  & 83.4 &      6.0 & 4 & 0.05 & 2518 & 0 & 0 & 44.5  & 83.4\\
6.0 & 4 & 0.05 & 2512 & 0 & 0 & 44.5  & 57.1 &      6.0 & 4 & 0.05 & 2521 & 0 & 0 & 44.5  & 76.2\\
6.0 & 4 & 0.05 & 2521 & 0 & 0 & 44.5  & 94.4 &      6.0 & 4 & 0.05 & 2510 & 1 & 0 & 58.1  & 106.3\\
6.0 & 4 & 1.0  & 3115 & 2 & 0 & 44.5  & 75.4 &      6.0 & 4 & 1.0  & 3121 & 2 & 0 & 80.4  & 124.5\\
6.0 & 4 & 1.0  & 3136 & 0 & 0 & 44.5  & 207.6 &     6.0 & 4 & 1.0  & 3127 & 2 & 0 & 44.5  & 116.5\\
6.0 & 4 & 1.0  & 3127 & 1 & 1 & 120.2 & 166.8 &     6.0 & 4 & 1.0  & 3115 & 3 & 0 & 44.7  & 81.1\\[1ex]
\enddata
\tablecomments{These simulations have a metallicity of $Z = 0.1 \, \zsun$. The left and right halves of the table are for simulations with the CE2 and CE0 collision prescriptions, respectively. The columns in the table for each simulation include the initial Elson profile slope ($\gamma$), virial radius ($\rv$), high-mass binary fraction ($\fbh$); total number of BHs formed through stellar collapse ($\nbh$), total number of BHs formed in the mass range $44.5 - 120 \, \msun$ ($N_{\rm PI}$), total number of IMBHs ($N_{\rm IM}$), the most massive BH ($M_{\rm BH, max}$), and the most massive star ($M_{\rm *, max}$). Rows marked with an asterisk (*) next to $M_{\rm BH, max}$ and $M_{\rm *, max}$ denote that a star more massive than $4000.0 \, \msun$ formed during the simulation, leading to its termination before $50\,\myr$. Simulations that terminate prematurely due to terminal energy error (see text) are denoted by $\dag$ or $\dag\dag$ symbols next to the values in $M_{\rm BH, max}$ and $M_{\rm *, max}$ columns. Simulations which stop prematurely but have the most massive star collapsing to form a black hole are marked with $\dag$. For simulations ending prematurely due to energy error but with the most massive star yet to collapse, we use the $\dag\dag$ symbol. In cases marked with * or $\dag\dag$, we take the instantaneous properties of the most massive star at the time of simulation termination as initial conditions and evolve it in a standalone fashion (no dynamics) in $\cosmic$ up to $50\,\myr$. In these cases, massive stars at the time of premature simulation termination are in core helium burning phase and that is why upon further standalone evolution using $\cosmic$ they quickly collapse converting most of their mass into an IMBHs.}
\end{deluxetable*}

\vspace{-1.2cm}
\startlongtable
\begin{deluxetable*}{@{\extracolsep{\fill}}lccccccc|lccccccr@{}}
\tabletypesize{\small}
\tablecaption{Relevant simulation parameters and outputs for \gridb\  cluster simulations with $\ncluster = 5 \times 10^5$ objects.} \label{tab:n5e5}


\tablehead{
 \colhead{$\gamma$} & \colhead{$\rv$} & \colhead{$\fbh$} &
 \colhead{$\nbh$} & \colhead{$N_{\rm PI}$} & \colhead{$N_{\rm IM}$} & 
 \colhead{$M_{\rm BH, max}$} & \colhead{$M_{\rm *, max}$} & 
 \colhead{ $\gamma$} & \colhead{$\rv$} & \colhead{$\fbh$} &
 \colhead{$\nbh$} & \colhead{$N_{\rm PI}$} & \colhead{$N_{\rm IM}$} & 
 \colhead{$M_{\rm BH, max}$} & \colhead{$M_{\rm *, max}$} \\[-1ex]
 \colhead{} & \colhead{($\pc$)} & \colhead{} &
 \colhead{} & \colhead{} & \colhead{} & 
 \colhead{($\msun$)} & \colhead{($\msun$)}  & 
 \colhead{} & \colhead{($\pc$)} & \colhead{} &
 \colhead{} & \colhead{} & \colhead{} & 
 \colhead{($\msun$)} & \colhead{($\msun$)}\\
\hline
\multicolumn{8}{c|}{\rule[10pt]{0pt}{0pt} \small{CE2}} & \multicolumn{8}{c}{\rule[15pt]{0pt}{0pt} \small{CE0}}
 }
\startdata
& & & & & & & & & & & & &\\ [-3ex]
2.3 & 1 & 0.05 & 1011 & 1 & 0 & 46.7  & 362.5 &     2.3 & 1 & 0.05 & 1031 &  2 & 1 & $1063.6^{\dag}$ & $1080.4^{\dag}$\\
2.3 & 1 & 0.05 & 999  & 0 & 0 & 44.5  & 324.7 &     2.3 & 1 & 0.05 & 1053 &  0 & 1 & 925.6 & 1024.5\\
2.3 & 1 & 0.05 & 1016 & 1 & 0 & 83.8  & 412.8 &     2.3 & 1 & 0.05 & 159  &  0 & 1 & $1854.3^{\dag\dag}$ & $1898.7^{\dag\dag}$\\
2.3 & 1 & 1.0  & 1442 & 6 & 0 & 84.9  & 468.3 &     2.3 & 1 & 1.0  & 624  &  0 & 2 & $2378.8^{\dag\dag}$ & $2422.6^{\dag\dag}$\\
2.3 & 1 & 1.0  & 1348 & 6 & 1 & 136.8 & 380.1 &     2.3 & 1 & 1.0  & 241  &  0 & 1 & $2345.5^{\dag\dag}$ & $2389.3^{\dag\dag}$\\
2.3 & 1 & 1.0  & 1430 & 4 & 1 & 148.2 & 450.6 &     2.3 & 1 & 1.0  & 108  &  1 & 1 & $2754.6^{\dag\dag}$ & $2798.4^{\dag\dag}$\\
3.0 & 1 & 0.05 & 1240 & 0 & 0 & 44.5  & 223.8 &     3.0 & 1 & 0.05 & 1205 &  0 & 1 & 1103.0 & 1123.2\\
3.0 & 1 & 0.05 & 1230 & 0 & 1 & 162.1 & 245.7 &     3.0 & 1 & 0.05 & 261  &  1 & 1 & $2244.9^{\dag\dag}$ & $2288.1^{\dag\dag}$\\
3.0 & 1 & 0.05 & 1235 & 2 & 1 & 225.0 & 392.3 &     3.0 & 1 & 0.05 & 1239 &  0 & 1 & 799.0 & 811.9\\
3.0 & 1 & 1.0  & 1829 & 6 & 1 & 154.3 & 405.9 &     3.0 & 1 & 1.0  & 1860 &  3 & 2 & 849.8 & 876.1\\
3.0 & 1 & 1.0  & 1797 & 8 & 3 & 149.5 & 327.0 &     3.0 & 1 & 1.0  & 1820 &  5 & 2 & 410.6 & 722.0\\
3.0 & 1 & 1.0  & 1778 & 5 & 1 & 124.6 & 369.5 &     3.0 & 1 & 1.0  & 1550 &  4 & 2 & $1507.1^{\dag}$ & $1510.0^{\dag}$\\
4.0 & 1 & 0.05 & 1287 & 1 & 0 & 97.8  & 141.0 &     4.0 & 1 & 0.05 & 1280 &  2 & 0 & 104.4 & 157.8\\
4.0 & 1 & 0.05 & 1282 & 1 & 0 & 101.9 & 137.5 &     4.0 & 1 & 0.05 & 1282 &  0 & 0 & 44.5 & 145.8\\
4.0 & 1 & 0.05 & 1277 & 1 & 0 & 51.1  & 177.9 &     4.0 & 1 & 0.05 & 1283 &  0 & 0 & 44.5 & 128.8\\
4.0 & 1 & 1.0  & 2070 & 6 & 1 & 184.0 & 231.5 &     4.0 & 1 & 1.0  & 2090 &  7 & 0 & 95.7 & 296.1\\
4.0 & 1 & 1.0  & 2047 & 5 & 0 & 68.0  & 241.6 &     4.0 & 1 & 1.0  & 2079 &  6 & 0 & 96.8 & 295.0\\
4.0 & 1 & 1.0  & 2098 & 6 & 0 & 104.1 & 317.1 &     4.0 & 1 & 1.0  & 2044 &  4 & 2 & 179.3 & 298.5\\
6.0 & 1 & 0.05 & 1329 & 0 & 0 & 44.5  & 140.3 &     6.0 & 1 & 0.05 & 1324 &  0 & 0 & 44.5 & 139.9\\
6.0 & 1 & 0.05 & 1321 & 0 & 0 & 44.5  & 104.2 &     6.0 & 1 & 0.05 & 1327 &  0 & 0 & 44.5 & 99.6\\
6.0 & 1 & 0.05 & 1323 & 0 & 0 & 44.5  & 136.9 &     6.0 & 1 & 0.05 & 1323 &  1 & 0 & 46.2 & 136.9\\
6.0 & 1 & 1.0  & 2089 & 0 & 0 & 44.5  & 223.3 &     6.0 & 1 & 1.0  & 2090 &  3 & 1 & 295.1 & 328.4\\
6.0 & 1 & 1.0  & 2090 & 3 & 0 & 104.2 & 190.1 &     6.0 & 1 & 1.0  & 2131 &  7 & 1 & 173.7 & 227.6\\
6.0 & 1 & 1.0  & 2070 & 2 & 0 & 76.8  & 233.8 &     6.0 & 1 & 1.0  & 2113 &  8 & 1 & 184.7 & 214.1\\[1ex]
\hline
& & & & & & & & & & & & &\\[-3ex] 
2.3 & 2 & 0.05 & 1099 & 0 & 0 & 44.5  & 452.6 &     2.3 & 2 & 0.05 & 1101 & 0 & 1 & 328.2 & 482.2\\
2.3 & 2 & 0.05 & 1122 & 1 & 1 & 167.7 & 477.0 &     2.3 & 2 & 0.05 & 1097 & 0 & 1 & 450.6 & 459.3\\
2.3 & 2 & 0.05 & 1095 & 0 & 1 & 135.5 & 270.3 &     2.3 & 2 & 0.05 & 1125 & 0 & 0 & 44.5 & 622.0\\
2.3 & 2 & 1.0  & 1502 & 3 & 0 & 93.1  & 360.6 &     2.3 & 2 & 1.0  & 1541 & 1 & 1 & 458.9 & 518.6\\
2.3 & 2 & 1.0  & 1491 & 6 & 0 & 58.7  & 264.4 &     2.3 & 2 & 1.0  & 1543 & 1 & 0 & 112.0 & 449.1\\
2.3 & 2 & 1.0  & 1502 & 1 & 0 & 116.4 & 268.9 &     2.3 & 2 & 1.0  & 1551 & 1 & 1 & 137.8 & 486.5\\
3.0 & 2 & 0.05 & 1268 & 0 & 0 & 44.5  & 94.2 &      3.0 & 2 & 0.05 & 1257 & 0 & 0 & 44.5 & 112.3\\
3.0 & 2 & 0.05 & 1262 & 0 & 0 & 44.5  & 92.3 &      3.0 & 2 & 0.05 & 1262 & 0 & 0 & 44.5 & 103.7\\
3.0 & 2 & 0.05 & 1271 & 0 & 0 & 44.5  & 101.0 &     3.0 & 2 & 0.05 & 1267 & 0 & 0 & 44.5 & 93.8\\
3.0 & 2 & 1.0  & 1727 & 0 & 0 & 44.5  & 155.4 &     3.0 & 2 & 1.0  & 1777 & 0 & 0 & 44.5 & 140.8\\
3.0 & 2 & 1.0  & 1810 & 2 & 0 & 72.9  & 178.1 &     3.0 & 2 & 1.0  & 1817 & 2 & 0 & 90.0 & 216.8\\
3.0 & 2 & 1.0  & 1814 & 1 & 0 & 85.7  & 215.8 &     3.0 & 2 & 1.0  & 1786 & 2 & 0 & 118.7 & 173.1\\\
4.0 & 2 & 0.05 & 1276 & 0 & 0 & 44.5  & 82.7 &      4.0 & 2 & 0.05 & 1276 & 0 & 0 & 44.5 & 88.3\\
4.0 & 2 & 0.05 & 1276 & 0 & 0 & 44.5  & 90.5 &      4.0 & 2 & 0.05 & 1285 & 0 & 0 & 44.5 & 90.5\\
4.0 & 2 & 0.05 & 1273 & 0 & 0 & 44.5  & 93.0 &      4.0 & 2 & 0.05 & 1271 & 0 & 0 & 44.5 & 93.0\\
4.0 & 2 & 1.0  & 1796 & 0 & 0 & 44.5  & 92.8 &      4.0 & 2 & 1.0  & 1749 & 0 & 0 & 44.5 & 155.3\\
4.0 & 2 & 1.0  & 1762 & 0 & 0 & 44.5  & 116.5 &     4.0 & 2 & 1.0  & 1790 & 0 & 0 & 44.5 & 124.6\\
4.0 & 2 & 1.0  & 1738 & 0 & 0 & 44.5  & 160.0 &     4.0 & 2 & 1.0  & 1780 & 1 & 0 & 45.6 & 160.0\\
6.0 & 2 & 0.05 & 1324 & 0 & 0 & 44.5  & 77.1 &      6.0 & 2 & 0.05 & 1319 & 0 & 0 & 44.5 & 130.9\\
6.0 & 2 & 0.05 & 1319 & 0 & 0 & 44.5  & 84.4 &      6.0 & 2 & 0.05 & 1317 & 0 & 0 & 44.5 & 65.5\\
6.0 & 2 & 0.05 & 1313 & 0 & 0 & 44.5  & 122.9 &     6.0 & 2 & 0.05 & 1321 & 0 & 0 & 44.5 & 122.9\\
6.0 & 2 & 1.0  & 1744 & 0 & 0 & 44.5  & 90.2 &      6.0 & 2 & 1.0  & 1762 & 0 & 0 & 44.5 & 90.6\\
6.0 & 2 & 1.0  & 1748 & 1 & 0 & 49.4  & 112.0 &     6.0 & 2 & 1.0  & 1765 & 0 & 0 & 44.5 & 162.3\\
6.0 & 2 & 1.0  & 1754 & 0 & 0 & 44.5  & 130.1 &     6.0 & 2 & 1.0  & 1737 & 0 & 0 & 44.5 & 130.1\\[1ex]
    \hline
& & & & & & & & & & & & &\\[-3ex] 
2.3 & 4 & 0.05 & 1089 & 0 & 0 & 44.5  & 136.6 &     2.3 & 4 & 0.05 & 1090 &  0 & 0 & 44.5  & 162.7\\
2.3 & 4 & 0.05 & 1099 & 0 & 0 & 44.5  & 99.0 &      2.3 & 4 & 0.05 & 1094 &  0 & 0 & 44.5  & 95.3\\
2.3 & 4 & 0.05 & 1091 & 0 & 0 & 44.5  & 89.9 &      2.3 & 4 & 0.05 & 1088 &  0 & 0 & 44.5  & 141.2\\
2.3 & 4 & 1.0  & 1395 & 1 & 0 & 45.9  & 283.5 &     2.3 & 4 & 1.0  & 1413 &  3 & 0 & 94.0  & 233.7\\
2.3 & 4 & 1.0  & 1431 & 0 & 1 & 128.2 & 204.3 &     2.3 & 4 & 1.0  & 1400 &  0 & 0 & 44.5  & 281.4\\
2.3 & 4 & 1.0  & 1432 & 3 & 0 & 65.9  & 266.8 &     2.3 & 4 & 1.0  & 1518 &  2 & 1 & 298.4 & 343.8\\
3.0 & 4 & 0.05 & 1229 & 0 & 0 & 44.5  & 78.0 &      3.0 & 4 & 0.05 & 1230 &  0 & 0 & 44.5  & 80.5\\
3.0 & 4 & 0.05 & 1232 & 0 & 0 & 44.5  & 58.7 &      3.0 & 4 & 0.05 & 1228 &  0 & 0 & 44.5  & 86.4\\
3.0 & 4 & 0.05 & 1232 & 0 & 0 & 44.5  & 91.6 &      3.0 & 4 & 0.05 & 1227 &  0 & 0 & 44.5  & 91.6\\
3.0 & 4 & 1.0  & 1460 & 0 & 0 & 44.5  & 99.5 &      3.0 & 4 & 1.0  & 1468 &  0 & 0 & 44.5  & 70.1\\
3.0 & 4 & 1.0  & 1459 & 0 & 0 & 44.5  & 84.1 &      3.0 & 4 & 1.0  & 1486 &  0 & 0 & 44.5  & 82.9\\
3.0 & 4 & 1.0  & 1495 & 0 & 0 & 44.5  & 113.8 &     3.0 & 4 & 1.0  & 1453 &  0 & 0 & 44.5  & 96.0\\
4.0 & 4 & 0.05 & 1270 & 1 & 0 & 44.5  & 89.7 &      4.0 & 4 & 0.05 & 1271 &  0 & 0 & 44.5  & 69.4\\
4.0 & 4 & 0.05 & 1273 & 0 & 0 & 44.5  & 42.7 &      4.0 & 4 & 0.05 & 1272 &  0 & 0 & 44.5  & 57.0\\
4.0 & 4 & 0.05 & 1269 & 0 & 0 & 44.5  & 44.9 &      4.0 & 4 & 0.05 & 1274 &  0 & 0 & 44.5  & 70.6\\
4.0 & 4 & 1.0  & 1552 & 0 & 0 & 44.5  & 94.6 &      4.0 & 4 & 1.0  & 1572 &  0 & 0 & 44.5  & 94.6\\
4.0 & 4 & 1.0  & 1554 & 3 & 0 & 44.5  & 64.4 &      4.0 & 4 & 1.0  & 1574 &  3 & 0 & 44.5  & 64.4\\
4.0 & 4 & 1.0  & 1562 & 2 & 0 & 44.5  & 105.2 &     4.0 & 4 & 1.0  & 1553 &  1 & 0 & 44.5  & 59.5\\
6.0 & 4 & 0.05 & 1311 & 0 & 0 & 44.5  & 79.8 &      6.0 & 4 & 0.05 & 1311 &  1 & 0 & 44.5  & 24.3\\
6.0 & 4 & 0.05 & 1307 & 0 & 0 & 44.5  & 55.4 &      6.0 & 4 & 0.05 & 1311 &  0 & 0 & 44.5  & 56.2\\
6.0 & 4 & 0.05 & 1313 & 0 & 0 & 44.5  & 29.6 &      6.0 & 4 & 0.05 & 1313 &  0 & 0 & 44.5  & 29.6\\
6.0 & 4 & 1.0  & 1570 & 2 & 0 & 44.5  & 57.3 &      6.0 & 4 & 1.0  & 1574 &  3 & 0 & 44.5  & 57.2\\
6.0 & 4 & 1.0  & 1552 & 1 & 0 & 44.5  & 50.8 &      6.0 & 4 & 1.0  & 1560 &  1 & 0 & 44.5  & 67.3\\
6.0 & 4 & 1.0  & 1555 & 1 & 0 & 44.5  & 80.8 &      6.0 & 4 & 1.0  & 1536 &  1 & 0 & 44.5  & 80.8\\[1ex]
\enddata
\tablecomments{Please refer to the note of Table~\ref{tab:n1e6} for the interpretation of symbols such as $\dag$ and $\dag\dag$ in columns $M_{\rm BH, max}$ and $M_{\rm *, max}$.}
\end{deluxetable*}
\vspace{-1.0cm}
\bibliography{main}
\bibliographystyle{aasjournal}




\end{document}